\documentclass[aps,prb,reprint,showpacs]{revtex4-1}
\usepackage[dvipdfmx]{graphicx}
\usepackage{subfigmat}
\usepackage{amsmath,amssymb}
\usepackage{color}
\usepackage{bm}
\usepackage{hyperref}
\usepackage[compact]{titlesec}
    \titlespacing{\section}{0pt}{2ex}{1ex}
    \titlespacing{\subsection}{0pt}{1ex}{0ex}
    \titlespacing{\subsubsection}{0pt}{0.5ex}{0ex}

\usepackage{cancel}

\usepackage{upgreek}

\newcommand{\tempinv}{\upbeta}

\begin{document}

\title{Spin pumping of two-dimensional electron gas with Rashba and Dresselhaus spin-orbit interactions}

\author{M. Yama$^{1}$, M. Tatsuno$^{1}$, T. Kato$^{1}$, and M. Matsuo$^{2,3,4,5}$}
\affiliation{
${^1}$Institute for Solid State Physics, The University of Tokyo, Kashiwa 277-8581, Japan\\
${^2}$Kavli Institute for Theoretical Sciences, University of Chinese Academy of Sciences, Beijing 100190, China\\
${^3}$CAS Center for Excellence in Topological Quantum Computation, University of Chinese Academy of Sciences, Beijing 100190, China\\
${^4}$Advanced Science Research Center, Japan Atomic Energy Agency, Tokai 319-1195, Japan\\
${^5}$RIKEN Center for Emergent Matter Science (CEMS), Wako, Saitama 351-0198, Japan\\
}

\date{\today}

\begin{abstract}
We theoretically consider spin pumping in a junction between a ferromagnetic insulator (FI) and a two-dimensional electron gas (2DEG) in which the Rashba and Dresselhaus spin-orbit interactions coexist. 
Using second-order perturbation theory, we derive an increase in linewidth in the case of an interfacial exchange coupling in a ferromagnetic resonance (FMR) experiment.
We clarify how the enhancement of Gilbert damping depends on the resonant frequency and spin orientation of the FI.
We show that this setup of an FMR experiment can provide information on the spin texture of 2DEG at the Fermi surface.
\end{abstract}
\maketitle 

\section{INTRODUCTION}
\label{sec:introduction}

Spin pumping has been studied intensively in the field of spintronics as a versatile method to generate spin current using magnetization dynamics~\cite{Tserkovnyak2002,Tserkovnyak2005}.
Spin pumping is used for injecting spins from a ferromagnet into an adjacent material in various heterojunction systems, such as ferromagnetic metal/normal metal (NM) junctions~\cite{Mizukami2001,Mizukami2002,Saitoh2006,Ando2008} and ferromagnetic insulator (FI)/NM junctions~\cite{Kajiwara2010}. 
As the backaction of the spin injection, the Gilbert damping of the ferromagnet is modulated.
Therefore, spin pumping can be used as a probe of nonequilibrium spin states of materials in the sense that information on spin transport due to an adjacent material is reflected in the modulation of the Gilbert damping.

An attractive strategy is spin pumping of semiconductor microstructures because highly developed semiconductor technologies can be utilized~\cite{Zutic2004,Awschalom2007}.
In particular, a two-dimensional electron gas (2DEG) in a semiconductor heterostructure is an easily controlled physical system that has been used in spintronics devices~\cite{Datta1990,Srisongmuang2008,Akabori2012,Feng2017}.
A 2DEG system has two types of spin{-}orbit interaction, Rashba~\cite{Bychkov1984,Rashba2015} and Dresselhaus ~\cite{Dresselhaus1955,Rocca1988}.
By combining these two interactions, we can control the spin polarization of conduction electrons in a way that is dependent on the propagation direction~\cite{Manchon2015}. 
In fact, electron transport reflecting the spin texture on the Fermi surface has been observed through, e.g., the Aharonov-Casher effect~\cite{Nitta1999,Ionicioiu2003,Frustaglia2004,Nitta2007,Nagasawa2013,Nagasawa2018} and the persistent spin helix state~\cite{Bernevig2006,Weber2007,Koralek2009,Sasaki2014,Schliemann2017,Iizasa2020,Zhao2020}.

In our work, we focus on spin pumping of a 2DEG in semiconductor heterostructures.
It is remarkable that there already exist experimental studies on injection of spins into two-dimensional electron systems with the Rashba spin-orbit interaction in surfaces ~\cite{Nakayama2016,Sanchez2013}, atomic layer materials~\cite{Ghiasi2019,Inoue2016}, transition oxides~\cite{Lesne2016,Song2017}, and bulk semiconductors~\cite{Ando2011,Sadovnikov2019}.
However, to the best of our knowledge, spin pumping in a junction system composed of an FI and a 2DEG in which the Rashba and Dresselhaus spin-orbit interactions coexist has not been studied yet. 

\begin{figure}[b]
\centering
\includegraphics[width=55mm]{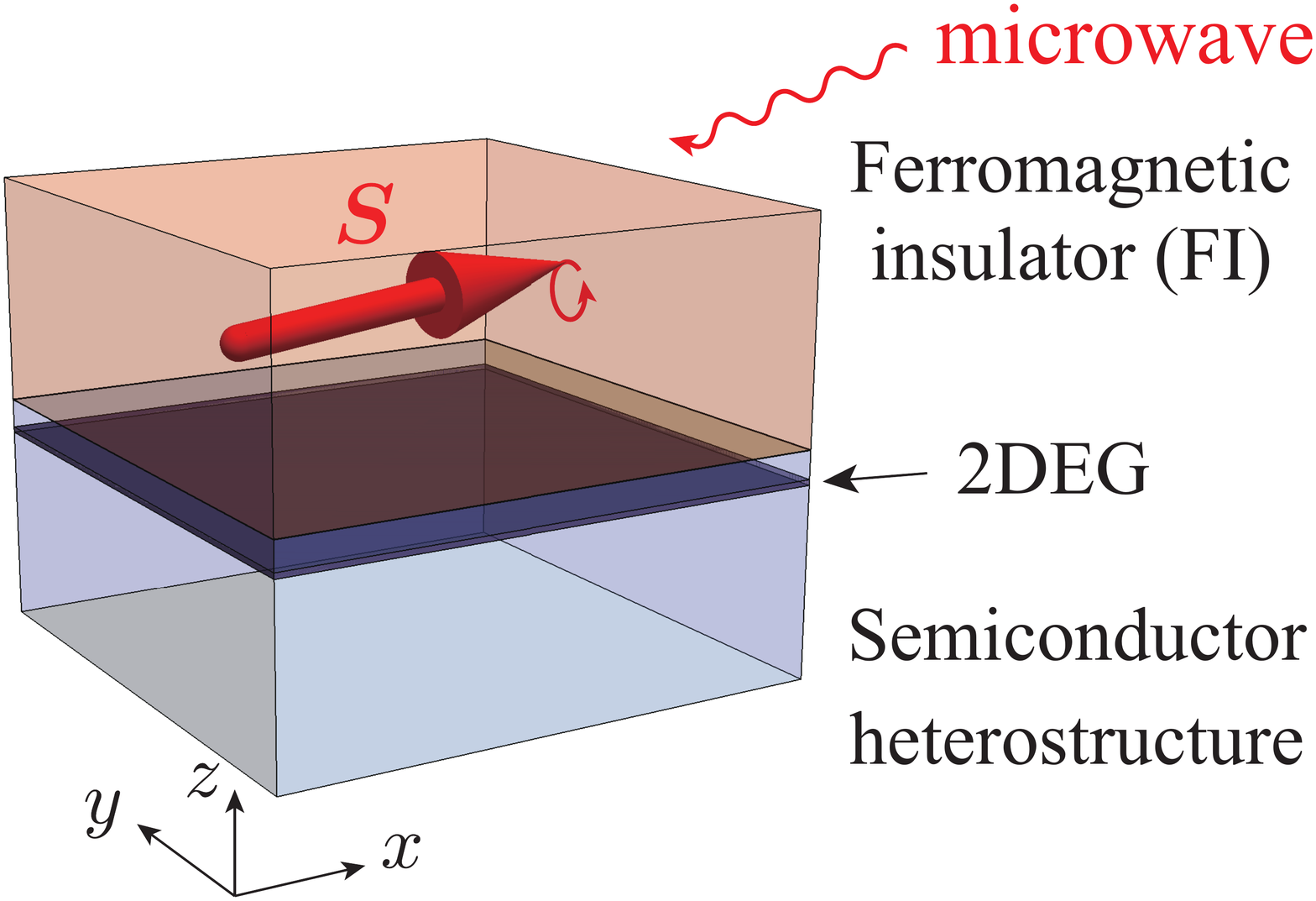}
\caption{Schematic picture of junction composed of a ferromagnetic insulator (FI) and two-dimensional electron gas (2DEG) in a semiconductor heterostructure.
Spin current is induced at the interface under microwave irradiation, whose frequency is chosen to be near the frequency of the spin precession of the FI.}
\label{fig:setup}
\end{figure}

In this study, we consider a planar junction composed of an FI and a 2DEG in a semiconductor heterostructure (see Fig.~\ref{fig:setup}).
By taking both Rashba and Dresselhaus spin{-}orbit interactions in 2DEG into account, we theoretically show an enhancement in Gilbert damping in the FI within the second-order perturbation with respect to interfacial exchange coupling~\cite{Ohnuma2014,Matsuo2018,Kato2019,Kato2020,Ominato2020a,Ominato2020b}.
We investigate how the linewidth increase in a ferromagnetic resonance (FMR) experiment depends on the resonant frequency and the spin orientation of the FI.
We show that a spin pumping measurement can directly detect the complex spin texture of a 2DEG with spin-orbit interactions.

The rest of this work is organized as follows.
In Sec.~\ref{sec:formulation}, we theoretically show an enhancement in Gilbert damping due to the exchange interaction with a 2DEG within second-order perturbation.
In Sec.~\ref{sec:result}, we show how the enhancement of the Gilbert damping depends on the spin orientation of the FI.
We also discuss the experimental relevance of our result.
Finally, we summarize our results in Sec.~\ref{sec:summary}.
In the subsequent four appendices, we give a detailed derivation of the equations in Sec.~\ref{sec:formulation}.

\section{FORMULATION}
\label{sec:formulation}

In this section, we analytically formulate the damping rate of the spin precession in a FMR experiment when the FI is coupled with a 2DEG through an interfacial interaction.
First, we introduce the model of the 2DEG and the FI in Sec.~\ref{sec:2DEGmodel} and Sec.~\ref{sec:FImodel}.
Next, we consider the interfacial coupling between the 2DEG and the FI in Sec.~\ref{sec:Interfacial}.
Finally, we calculate the second-order perturbation with respect to the interfacial coupling in Sec.~\ref{sec:SecondOrderPerturbation}.
Throughout this paper, we will use the laboratory coordinates shown in Fig.~\ref{fig:setup}; the $xy$ plane is parallel to the 2DEG, while the $z$-axis is perpendicular to the junction area.

\subsection{Two-dimensional electron gas}
\label{sec:2DEGmodel}

We consider a 2DEG with both Rashba and Dresselhaus interactions.
The Hamiltonian for the kinetic energy and the spin-orbit interactions is given as
\begin{align}
H_{\rm kin} &= \sum_{\bm k} (c_{{\bm k}\uparrow} ^\dagger \ c_{{\bm k}\downarrow}^\dagger) \hat{h}_{\bm k} \Bigl( \begin{array}{cc} c_{{\bm k}\uparrow} \\ c_{{\bm k}\downarrow} \end{array} \Bigr), \\
\hat{h}_{\bm k} &= \Bigl(\frac{\hbar^2(k_x^2+k_y^2)}{2m^*} -\mu\Bigr)
\hat{I} + \alpha(k_y \sigma_x -  k_x \sigma_y) \nonumber \\
& \hspace{10mm} + \beta (k_x \sigma_x - k_y \sigma_y), 
\label{eq:HamNM1}
\end{align}
where $\hat{h}_{\bm k}$ is a $2\times2$ matrix, $\sigma_a$ ($a=x,y,z$) are the Pauli matrices, $\hat{I}$ is an identity matrix, and $c_{{\bm k}\sigma}$ is the annihilation operator of conduction electrons with wavenumber ${\bm k}=(k_x,k_y)$ and $z$-component of a spin, $\sigma$ ($=\uparrow, \downarrow$).
The first term of $\hat{h}_{\bm k}$ describes the kinetic energy of an electron with chemical potential $\mu$ and effective mass $m^*$.
The second and third terms of $\hat{h}_{\bm k}$ describe the Rashba and Dresselhaus spin-orbit interactions, respectively, and $\alpha$ and $\beta$ denote the amplitudes of the respective spin-orbit interactions.

\begin{figure}[tb]
\centering
\includegraphics[width=\linewidth]{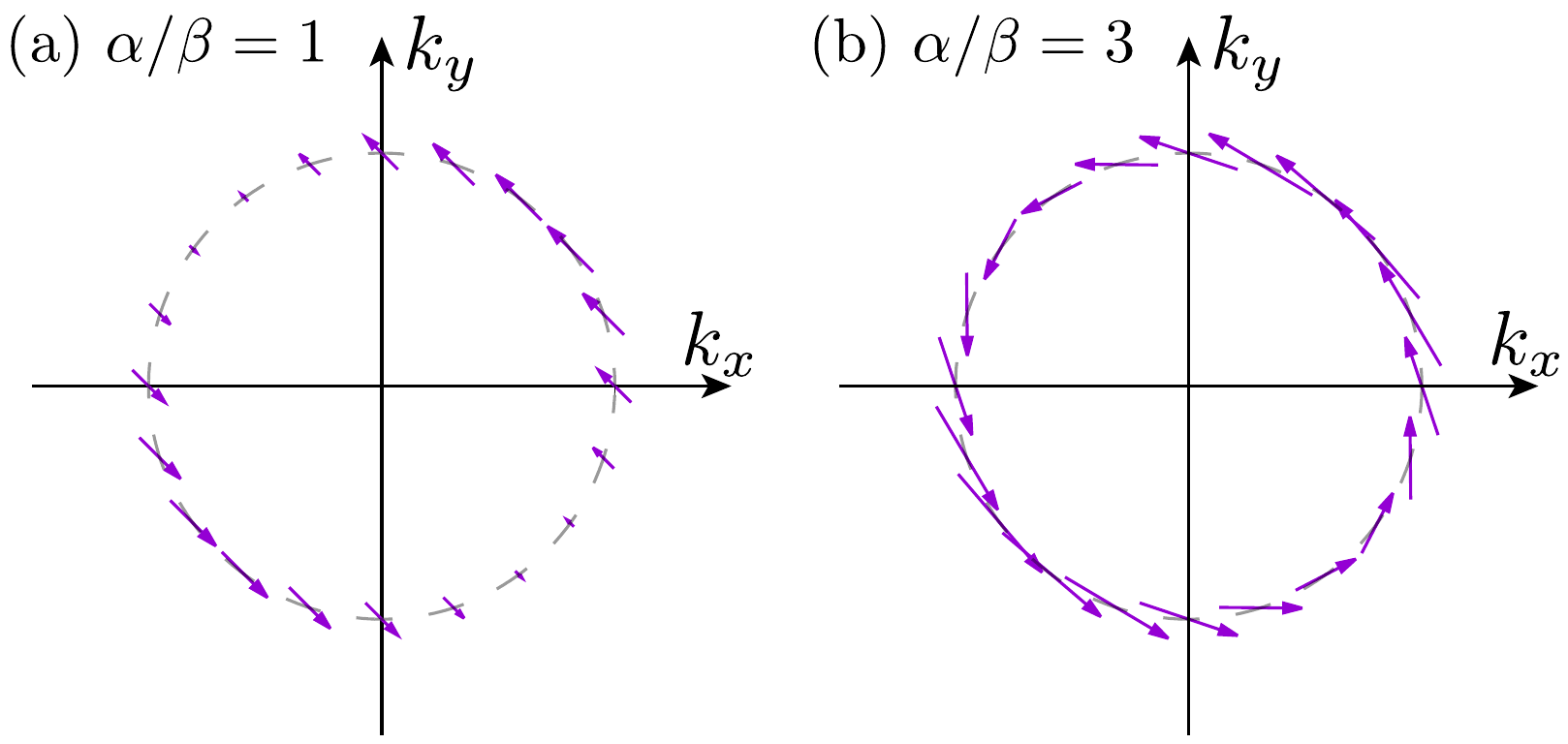}
\caption{Schematic picture of the effective magnetic field ${\bm h}_{\rm eff}(\varphi)$ that acts on the spins of conduction electrons.
}
\label{fig:spin_texture}
\end{figure}

Hereafter, we assume that the Fermi energy is much larger than the other energy scales, such as the temperature and the amplitude of the spin-orbit interactions.
The energy dispersion is then approximated as
$\xi_{\bm k}\equiv \hbar^2\bm k^{2}/2m^* - \mu \simeq v_{\rm F} (|{\bm k}|-k_{\rm F})$, where $k_{\rm F}$ and $v_{\rm F}$ are the Fermi wavenumber and the Fermi velocity, respectively.
We also approximate the spin-orbit interaction using $(k_x,k_y) \simeq (k_{\rm F}\cos \varphi,k_{\rm F} \sin \varphi)$, where $\varphi$ is the azimuth angle of the conduction electrons.
Accordingly, the Hamiltonian matrix can be rewritten as $\hat{h}_{\bm k} = \xi_{\bm k} \hat{I} - {\bm h}_{\rm eff}\cdot {\bm \sigma}$, where
\begin{align}
{\bm h}_{\rm eff} &\simeq k_{\rm F}(-\alpha \sin \varphi - \beta \cos \varphi, \alpha \cos \varphi + \beta \sin \varphi,0) ,
\end{align}
is the effective magnetic field that acts on the spin of a conduction electron propagating in the direction of $\varphi$.
Note that the effective magnetic field ${\bm h}_{\rm eff}$ characterizes the spin texture of conduction electrons as a function of the azimuth angle of the electron wavenumber.
Fig.~\ref{fig:spin_texture}~(a) and (b) shows profiles of the effective magnetic field on the Fermi surface{\footnote{{In Fig.~\ref{fig:spin_texture}, the spin-splitting of the Fermi surface has not been shown explicitly; it is assumed to be much smaller than the Fermi wavenumber.}}} for $\alpha/\beta=1$ and 3.
The spin texture is simplified for the special case of $\alpha/\beta=1$; the direction of the effective field is always parallel to the straight line $k_y = -k_x$.

By diagonalizing $\hat{h}_{\bm k}$, we obtain the spin-dependent electron dispersion,
\begin{align}
&E_{\bm k}^{\pm} = \xi_{\bm k} \pm h_{\rm eff}(\varphi), \\
& h_{\rm eff}(\varphi) \equiv |{\bm h}_{\rm eff}(\varphi)| \simeq
k_{\rm F}\sqrt{\alpha^2 + \beta^2 + 2\alpha\beta \sin 2 \varphi}.
\label{eq:heffdef}
\end{align}
Note that $2h_{\rm eff}(\varphi)$ corresponds to the spin-splitting energy for a conduction electron propagating at the azimuth angle $\varphi$.
When $\alpha=\beta$ ($\alpha=-\beta$), the spin-splitting energy vanishes at $\varphi=3\pi/4$ ($\varphi=\pi/4$).

Green's function at finite temperature is defined as a $2\times 2$ matrix $\hat{g}({\bm k},\tau)$, whose elements are
\begin{align}
g_{\sigma \sigma'}({\bm k},\tau) &= -\hbar^{-1} \langle c_{{\bm k}\sigma}(\tau) c_{{\bm k}\sigma'}^\dagger \rangle, 
\end{align}
for $0 < \tau < \hbar \tempinv$, where $c_{{\bm k}\sigma}(\tau) = e^{H_{\rm kin}\tau/\hbar} c_{{\bm k}\sigma} e^{-H_{\rm kin}\tau/\hbar}$ is the Heisenberg representation for the imaginary time evolution and $\tempinv$ is inverse temperature.
Note that there are off-diagonal components in the Green's function that are due to spin-flip processes by the spin-orbit interactions.
The Fourier transform of the Green's function is defined as
\begin{align}
\hat{g}({\bm k},i\omega_n) = \int_0^{\hbar \tempinv} d\tau \,  e^{i\omega_n\tau} \hat{g}({\bm k},\tau),
\end{align}
where $\omega_n=\pi (2n+1)/\hbar \tempinv$ are the Matsubara frequencies for fermions.
For the Hamiltonian $H_{\rm kin}$, the Green's function is calculated as
\begin{align}
&\hat{g}({\bm k},i\omega_n)= \frac{(i\hbar \omega_n-\xi_{\bm k})\hat{I} 
- {\bm h}_{\rm eff}\cdot {\bm \sigma}}
{(i\hbar\omega_n-E_{\bm k}^+)(i\hbar\omega_n-E_{\bm k}^-)}.
\end{align}

We also consider the effect of impurity scattering in a 2DEG by the Hamiltonian,
\begin{align}
V_{\rm imp} &= u \sum_{i\in {\rm imp}} \sum_{\sigma}{ {\Psi^\dagger({\bm R}_i) \Psi({\bm R}_i)}}
\\
{{\Psi({\bm r})}} &= \frac{1}{\sqrt{{{{\cal A}}}}} \sum_{\bm k} c_{{\bm k}\sigma} e^{i {\bm k}\cdot {{\bm r}}},
\end{align}
where $i\in {\rm imp}$ indicates the impurity site, $u$ is the strength of the impurity potential, {{${\cal A}$ is the junction area}}, and ${\bm R}_i$ is the position of the impurity site.
The total Hamiltonian for the 2DEG is thus $H_{\rm 2DEG} = H_{\rm kin} +V_{\rm imp}$.
Through second-order perturbation with respect to the impurity potential, Green's function is calculated as
\begin{align}
& \hat{g}({\bm k},i\omega_n) = \frac{(i \hbar \omega_n  -\xi_{\bm k}+i \Gamma {\rm sgn}(\omega_n)/2)\hat{I} - {\bm h}_{\rm eff}\cdot {\bm \sigma}}{\prod_{\nu=\pm} (i\hbar\omega_n-E_{\bm k}^\nu +i\Gamma {\rm sgn}(\omega_n)/2)},
\label{eq:gkwimp}
\end{align}
where $\Gamma = 2\pi n_i u^2 D(\epsilon_{\rm F}) $ is level broadening and $D(\epsilon_{\rm F}) = k_{\rm F}/(2\pi \hbar v_{\rm F})$ is the density of states per spin per unit volume.
For a detailed derivation, see Appendix~\ref{sec:ImpurityScattering}.

\subsection{Ferromagnetic insulator}
\label{sec:FImodel}
\begin{figure}[tb]
\centering
\includegraphics[width=40mm]{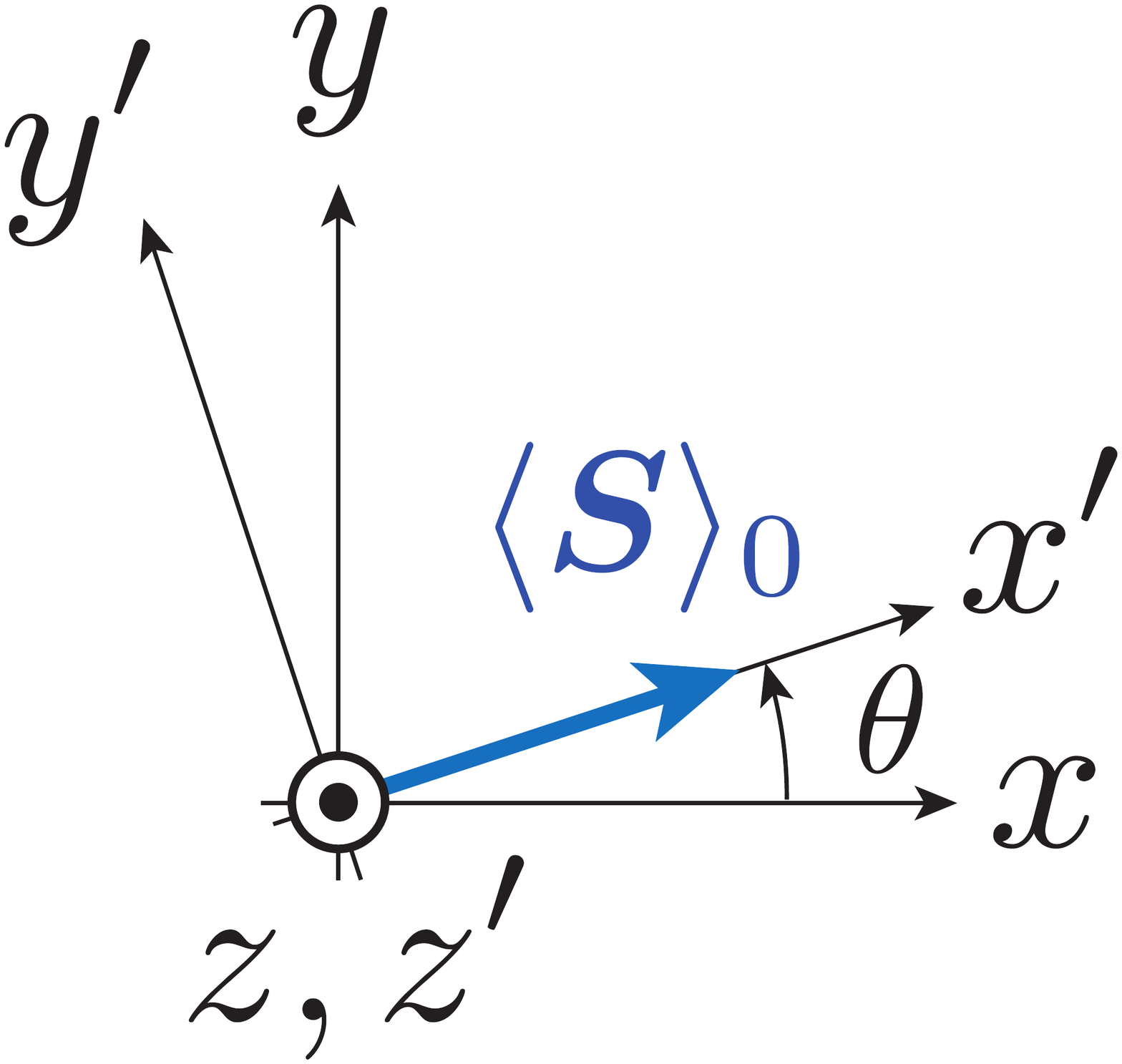}
\caption{Relation between laboratory coordinates $(x,y,z)$ and magnetization-fixed coordinates $(x',y',z')$ for FI.}
\label{fig:axis}
\end{figure}

We assume that the magnetization of the FI is in the $xy$ plane parallel to the 2DEG.
We define the direction of the in-plane ordered spin measured from the $x$-axis as $\theta$.
Accordingly, the expectation value of the ordered spin in FI is expressed as
\begin{align}
\langle {\bm S}_{i} \rangle = 
(\langle S_i^x \rangle,\langle S_i^y \rangle,\langle S_i^z \rangle)
= (S_0 \cos \theta, S_0 \sin \theta, 0),
\end{align}
where $i$ is the index of a localized spin site, and $S_0$ is the amplitude of the average spin per site.
To simplify the subsequent calculation, we introduce a new coordinate $(x',y',z')$, for which the direction of the ordered spin is fixed to the $x'$-axis, as shown in Fig.~\ref{fig:axis}.
The average spin is expressed in this new coordinate as $(\langle S_i^{x'} \rangle,\langle S_i^{y'} \rangle,\langle S_i^{z'} \rangle)
= (S_0, 0, 0)$.
The spin operators for the new coordinates are related to the ones for the original coordinates as
\begin{align}
S_i^{x'} &= \cos\theta S_i^{x} + \sin \theta S_i^{y}, \\
S_i^{y'} &= -\sin \theta S_i^{x} + \cos \theta S_i^{y}, \\
S_i^{z'} &= S_i^{z}.
\end{align}
The Hamiltonian of the FI is written in the new coordinates as
\begin{align}
H_{\rm {FI}}&=\sum_{\langle i,j \rangle}J_{ij} (S_{i}^{x'} S_{j}^{x'} + S_{i}^{y'} S_{j}^{y'} + S_{i}^{z'} S_{j}^{z'})  \nonumber \\
&-\hbar \gamma h_{\rm dc}\sum_{i}S_{i}^{x'},
\end{align}
where $\langle i,j \rangle$ indicates a pair of nearest-neighbor sites, $J_{ij}$ is the ferromagnetic exchange coupling, $\gamma$($<0$) is the gyromagnetic ratio, and $h_{\rm dc}$ is the static magnetic field.
We assume $h_{\rm dc} < 0$, for which the direction of the spin (the magnetization) becomes $+x'$ ($-x'$).

We further assume that the temperature is much lower than the magnetic transition temperature.
As well, we assume $S_{0}\gg 1$, to which the spin-wave approximation can be applied.
The Hamiltonian of the FI can then be rewritten as
\begin{align}
    H_{{\rm FI}} &= \sum_{\bm k} \hbar \omega_{\bm k} b_{\bm k}^\dagger b_{\bm k}, 
    \label{eq:HFI1}
    \\
\hbar \omega_{\bm k} &= {\cal D}{\bm k}^2 + \hbar \gamma h_{\rm dc},
\label{eq:HFI2}
\end{align}
where $b_{\bm k}$ is the annihilation operator of a magnon, and ${\cal D}$ is spin stiffness.
A detailed derivation is given in Appendix~\ref{app:SpinWave}.

We define the imaginary-time spin correlation function of the FI as
\begin{align}
& G({\bm k},\tau) 
 = -\frac{1}{\hbar} \langle 
    S^{x'+}_{{\bm k}}(\tau),S^{x'-}_{{\bm k}}(0)
    \rangle , \\
& G({\bm k},i\omega_n) = \int_{0}^{\hbar \tempinv} d\tau \, G({\bm k},\tau) e^{i\omega_n \tau},
\end{align}
where $\omega_n = 2\pi n/\hbar \tempinv$ are the Matsubara frequencies for bosons.
This spin correlation function can be calculated within the spin-wave approximation as
\begin{align}
G({\bm k},i\omega_n) & =\frac{2S_0/\hbar}{i\omega_n-\omega_{\bm k}+i\alpha_{\rm G} |\omega_n|}. \label{eq:correlation_magnon}
\end{align}
Here, we have introduced a phenomenological dimensionless parameter $\alpha_{\rm G}$ that describes the strength of the Gilbert damping.

\subsection{Interfacial exchange interaction}
\label{sec:Interfacial}

We consider an interfacial exchange interaction between the FI and the 2DEG with the Hamiltonian,
\begin{align}
H_{\rm int} &= \sum_{{\bm k}} ({\cal T}_{\bm k} S_{\bm k}^{x'+} s_{\bm k}^{x'-}
+ {\cal T}_{\bm k}^* S_{\bm k}^{x'-} s_{\bm k}^{x'+}).
\end{align}
where ${\cal T}_{\bm k}$ is an exchange interaction at the interface, and $S_{\bm k}^{x'\pm} = S_{\bm k}^{y'} \pm i S_{\bm k}^{z'}$ are creation and annihilation operators of the localized spins in the FI.
Here, the interface is assumed to be sufficiently flat so that the momentum of spin excitation is conserved.
{The effective field (the exchange bias) which acts on the conduction electrons via the interfacial coupling is also assumed to be much smaller than the effective magnetic field ${\bm h}_{\rm eff}$.}
The spin ladder operators for conduction electrons, $s_{\bm k}^{x'\pm}$, are defined as follows.
First, we define the spin operators in the magnetization-fixed coordinates as
\begin{align}
s_{\bm k}^{x'} &= \cos \theta \, s_{\bm k}^{x}
+ \sin \theta \, s_{\bm k}^{y}, \label{eq:start} \\
s_{\bm k}^{y'} &= -\sin \theta \, s_{\bm k}^{x}
+ \cos \theta \, s_{\bm k}^{y}, \\
s_{\bm k}^{z'} &= s_{\bm k}^{z},
\end{align}
using the spin operators in the original coordinates defined as
\begin{align}
s^{a}_{\bm k} = \sum_{\sigma \sigma'} \sum_{{\bm k}'} c_{{\bm k}'\sigma}^\dagger (\sigma_{a})_{\sigma \sigma'} c_{{\bm k}'+{\bm k}\sigma}, \quad (a=x,y,z) .
\label{eq:end} 
\end{align}
Here, $\sigma_a$ are the Pauli matrices.
We define the spin ladder operators as
\begin{align}
s_{\bm k}^{x'+} &\equiv s_{\bm k}^{y'} + i s_{\bm k}^{z'}, \\
s_{\bm k}^{x'-} &\equiv (s_{{\bm k}}^{x'+})^\dagger
= s_{-{\bm k}}^{y'} - i s_{-{\bm k}}^{z'}.
\end{align}
Using Eqs.~(\ref{eq:start})-(\ref{eq:end}), these ladder operators are rewritten as
\begin{align}
s_{\bm k}^{x'\pm} &=
\frac12 \sum_{\sigma,\sigma'} \sum_{{\bm k}'} c_{{\bm k}'\sigma}^\dagger
(\hat{\sigma}^{x'\pm})_{\sigma\sigma'} c_{{\bm k}'\pm {\bm k}\sigma'},
\label{eq:skpm} \\
\hat{\sigma}^{x'\pm} &= -\sin \theta \, \sigma_x + \cos \theta \, \sigma_y \pm i \sigma_z.
\label{eq:skmmatrix}
\end{align}

\subsection{Second-order perturbation}
\label{sec:SecondOrderPerturbation}

Let us consider a second-order perturbation with respect to the interfacial exchange interaction $H_{\rm int}$ to the bulk system described by $H_{\rm 2DEG}+H_{\rm FI}$.
The spin correlation function is written in the form,
\begin{align}
G({\bm k},i\omega_{n}) &= \frac{1}{(G_0({\bm k},i\omega_n))^{-1}-\Sigma({\bm k},i\omega_n)}, \\
\Sigma({\bm k},i\omega_n) 
&= \frac{|T_{\bm k}|^2}{4\tempinv}
\sum_{{\bm k}',i\omega_m} {\rm Tr} \Biggl[ \hat{\sigma}^{x'-}
\hat{g}({\bm k}',i\omega_m) \nonumber \\
& \hspace{7mm} \times \hat{\sigma}^{x'+}
\hat{g}({\bm k}'+{\bm k},i\omega_m+i\omega_n) \Biggr]\label{eq:self_Sigma},
\end{align}
where $G_0({\bm k},i\omega_n)$ is the unperturbed part of the Green's function given in Eq.~(\ref{eq:correlation_magnon}), and $\Sigma({\bm k},i\omega_n)$ is the self-energy due to the interfacial exchange coupling.
Since a uniform spin precession is induced in FMR experiments, we only calculate the self-energy for ${\bm k}={\bm 0}$.
To simplify the notation, we will rewrite Green's function $\hat{g}({\bm k},i\omega_n)$ as
\begin{align}
\hat{g}(\bm k,i\omega_{n}) &= \frac{A(i\omega_{n})\hat{I}-{\bm h}_{\rm eff}\cdot {\bm \sigma}
}{D(i\omega_{n})}, \\
A(i\omega_{n}) &= i\hbar \omega_{n}-\xi_{\bm k}+i\Gamma{\rm sgn}(\omega_{n})/2, \label{eq:Adef} \\
D(i\omega_{n}) &= \prod_{\nu = \pm} (i\hbar \omega_{n}-E^{\nu}_{\bm k}+i\Gamma{\rm sgn}(\omega_{n})/2).
\label{eq:Ddef}
\end{align}
By a straightforward calculation using the algebra of the Pauli matrices, we obtain
\begin{align}
&\Sigma({\bm q}={\bm 0},i\omega_n) \nonumber \\
& = \frac{|{\cal T}_{\bm 0}|^2}{\tempinv}
\sum_{{\bm k},i\omega_m}
\frac{A - {\bm h}_{\rm eff} \cdot \hat{\bm m}}{D}
\frac{A' + {\bm h}_{\rm eff} \cdot \hat{\bm m}}{D'},
\label{eq:sigmaresult1}
\end{align}
where $\hat{\bm m} = (\cos \theta,\sin \theta,0)$ indicates the spin orientation of the FI, $A=A(i\omega_m)$, $A'=A(i\omega_m+i\omega_n)$, $D=D(i\omega_m)$, and $D'=D(i\omega_m+i\omega_n)$ (for the detailed derivation, see Appendix~\ref{app:CalcSpin}).
For further calculation, we evaluate the sum with respect to the Matsubara frequency $\omega_m$ by using the standard procedure based on the contour integral.
Then, we obtain the retarded component by analytic continuation: $i\omega_n \rightarrow \omega + i\delta$ (the details of the calculation are in Appendix~\ref{app:AnalyticContinuation}).

We will focus on the increase in the FMR linewidth due to the exchange coupling with the 2DEG and consider only the imaginary part of the self-energy.
Using the unperturbed spin correlation function Eq.~(\ref{eq:correlation_magnon}), we obtain the retarded component of the spin (magnon) correlation function as
\begin{align}
G^R({\bm q=\bm 0},\omega) 
&\simeq\frac{2S_{0}/\hbar}{\omega-\omega_{\bm q=\bm 0}+i(\alpha_{\rm G}+\delta\alpha_{\rm G})\omega},\label{eq:green_magnon}\\
\delta\alpha_{\rm G}(\omega) &\equiv -\frac{2S_{0}}{\hbar\omega}{\rm Im}\, \Sigma^R(\bm q=\bm 0,\omega).
\label{eq:deltagamma}
\end{align}
For $\alpha_{\rm G} +\delta \alpha_{\rm G} \ll 1$ {that holds for a standard setup of spin pumping,}
the linewidth of the ferromagnetic resonance is sufficiently small, allowing us to replace $\omega$ with the resonance frequency $\omega_{{\bm q}={\bm 0}}$ to express the enhancement of the damping constant in Eq.~(\ref{eq:deltagamma}).
Hereafter, the resonance frequency is simply written as $\Omega (\equiv \omega_{{\bm q}={\bm 0}})$.
Then, ${\rm Im}\, \Sigma^R(\bm q=\bm 0,\Omega)$ corresponds to the increase in the linewidth in the FMR experiment.
Analytic calculation of ${\rm Im}\, \Sigma(\bm q=\bm 0,\Omega)$ gives a correction for the Gilbert damping as
\begin{align}
\delta \alpha_{\rm G} (\Omega) 
&\simeq -\frac{2S_{0}}{\hbar\Omega}{\rm Im}\, \Sigma^{{R}}(\bm q=\bm 0,\Omega) \nonumber \\
& = \alpha_{{\rm G},0} \sum_{\nu,\nu'=\pm 1}\int_0^{2\pi} \! \frac{d\varphi}{2\pi} \,
F\bigl(\hbar \Omega+(\nu-\nu')h_{\rm eff} \bigr) \nonumber \\
& \hspace{10mm} \times \frac{1-\nu \hat{\bm h}_{\rm eff}(\varphi) \cdot \hat{\bm m}}{2}
\frac{1+\nu '\hat{\bm h}_{\rm eff}\cdot \hat{\bm m}}{2},
\label{eq:mainresult} \\
F(x) &= \frac{\Gamma/\pi \Delta_0}{(x/\Delta_0)^2+(\Gamma/\Delta_0)^2},
\end{align}
where $\varphi$ describes the propagation direction of conduction electrons, $\alpha_{{\rm G},0} = 2 \pi S_0|{\cal T}_0|^2 {\cal A} D(\epsilon_{\rm F})/\Delta_0$ is a dimensionless parameter, ${\cal A}$ is the
junction area, and $\hat{\bm h}_{\rm eff}(\varphi)={\bm h}_{\rm eff}(\varphi)/h_{\rm eff}(\varphi)$ is a unit direction vector of the effective magnetic field.
Here, we have introduced the unit of energy $\Delta_0$, which is the amplitude of the Dresselhaus spin-orbit interaction, $k_{\rm F} \beta$. 
The enhancement of the Gilbert damping can be separated into three parts:
\begin{align}
\delta\alpha_{\rm G} & = \delta \alpha_{{\rm G},1} + \delta \alpha_{{\rm G},2}+ \delta \alpha_{{\rm G},3} \\
\delta \alpha_{{\rm G},1} &= \alpha_{{\rm G},0} \int_0^{2\pi} \! \frac{d\varphi}{2\pi} \, F(\hbar \Omega) \frac{1-(\hat{\bm h}_{\rm eff}\cdot \hat{\bm m})^2}{2}, 
\label{eq:result1} \\
\delta \alpha_{{\rm G},2} &= \alpha_{{\rm G},0} \int_0^{2\pi} \! \frac{d\varphi}{2\pi} \, F(\hbar \Omega-2h_{\rm eff}) \frac{(1+\hat{\bm h}_{\rm eff}\cdot \hat{\bm m})^2}{4}, 
\label{eq:result2} \\
\delta \alpha_{{\rm G},3} &= \alpha_{{\rm G},0} \int_0^{2\pi} \! \frac{d\varphi}{2\pi} \, F(\hbar \Omega+2h_{\rm eff}) \frac{(1-\hat{\bm h}_{\rm eff}\cdot \hat{\bm m})^2}{4},
\label{eq:result3}
\end{align}
These expressions indicate the physical mechanism of the enhancement of the Gilbert damping as follows.
The first contribution $\delta \alpha_{{\rm G},1}$ comes from elastic spin flipping of conduction electrons caused by the transverse component of the effective magnetic field ${\bm h}_{\rm eff}$ via the interfacial exchange coupling.
In fact, $\delta \alpha_{{\rm G},1}$ vanishes when ${\bm h}_{\rm eff}$ is parallel or anti-parallel to the magnetization of FI, $\hat{\bm m}$.
Since this process is elastic, the frequency-dependent part is just a Lorentzian form $F(\hbar \Omega)$, which has a peak at $\Omega = 0$.
The second contribution $\delta \alpha_{{\rm G},2}$ originates from the magnon absorption process.
This is a dynamical process as indicated by the peak shift of the Lorentzian form; the peak of $F(\hbar \Omega-2h_{\rm eff}(\varphi))$ is shifted to $\Omega = 2h_{\rm eff}(\varphi)/\hbar$, at which the magnon energy coincides with the spin-splitting energy gap of conduction electrons.
The second contribution takes a maximum when $\hat{\bm h}_{\rm eff}$ is parallel to $\hat{\bm m}$.
This is consistent with the fact that a spin of a conduction electron is converted from a low-energy state $\hat{\bm h}_{\rm eff}$ to a higher state $-\hat{\bm h}_{\rm eff}$ by receiving a spin carried by a magnon, that is in the direction of $-\hat{\bm m}$.
We should also note that the second contribution vanishes when $\hat{\bm h}_{\rm eff}$ is anti-parallel to $\hat{\bm m}$.
The third contribution $\delta \alpha_{{\rm G},3}$ comes from the magnon emission process.
This process is usually not important in FMR experiments, because the frequency $\Omega$ is taken as positive.

\section{RESULTS}
\label{sec:result}

\subsection{Enhancement of Gilbert damping}
\begin{figure}[t]
\centering
\includegraphics[width=0.9\linewidth]{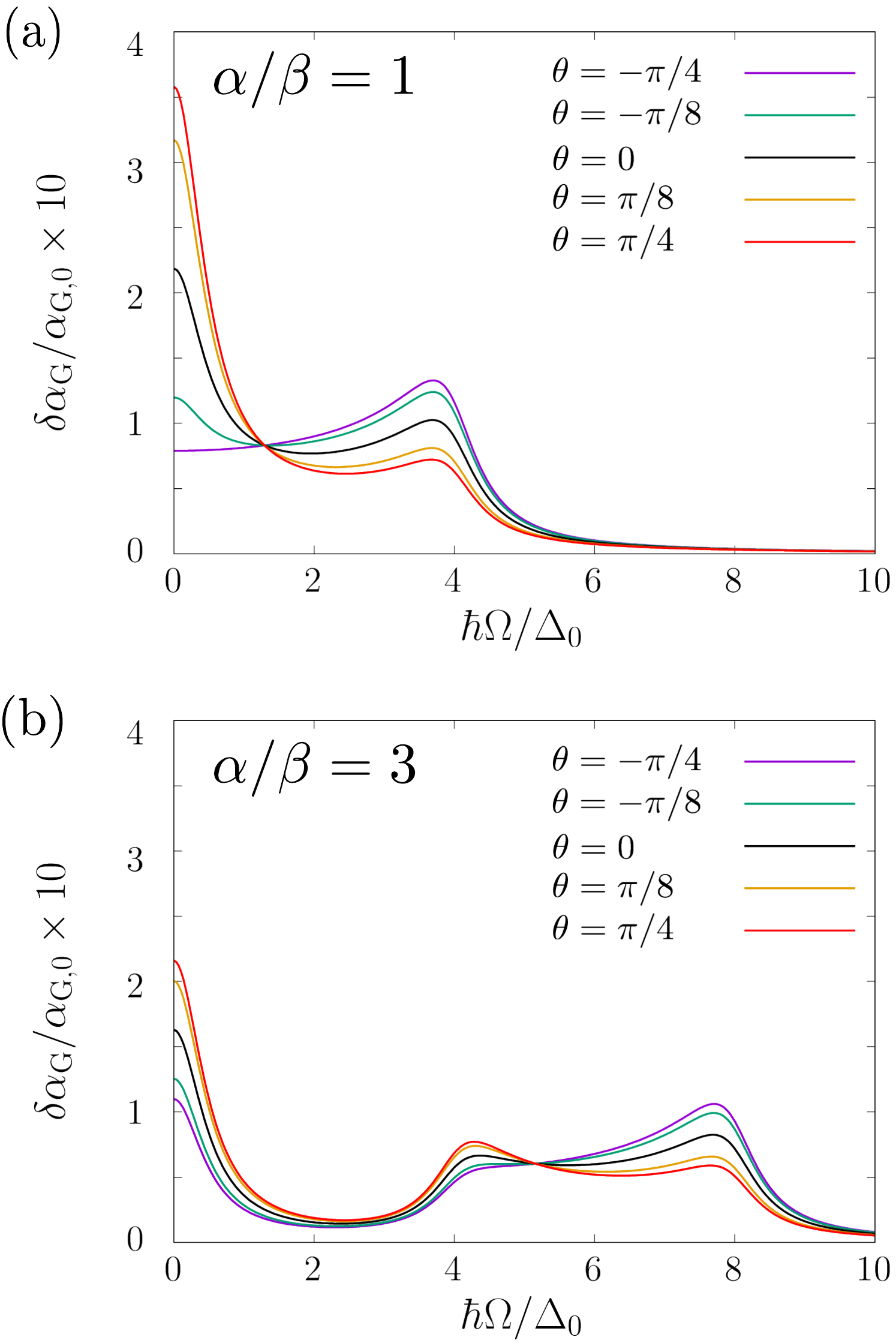}
\label{fig:Gilbert}
\caption{Enhancement of Gilbert damping, $\delta\alpha_{\rm G}$, due to coupling to 2DEG plotted as a function of FMR resonance frequency $\Omega$. (a) $\alpha/\beta = 1$. (b) $\alpha/\beta = 3$.}
\label{fig:increase_Gilbert_correction_term}
\end{figure}

\begin{figure*}[tb]
\centering
\includegraphics[width = \linewidth]{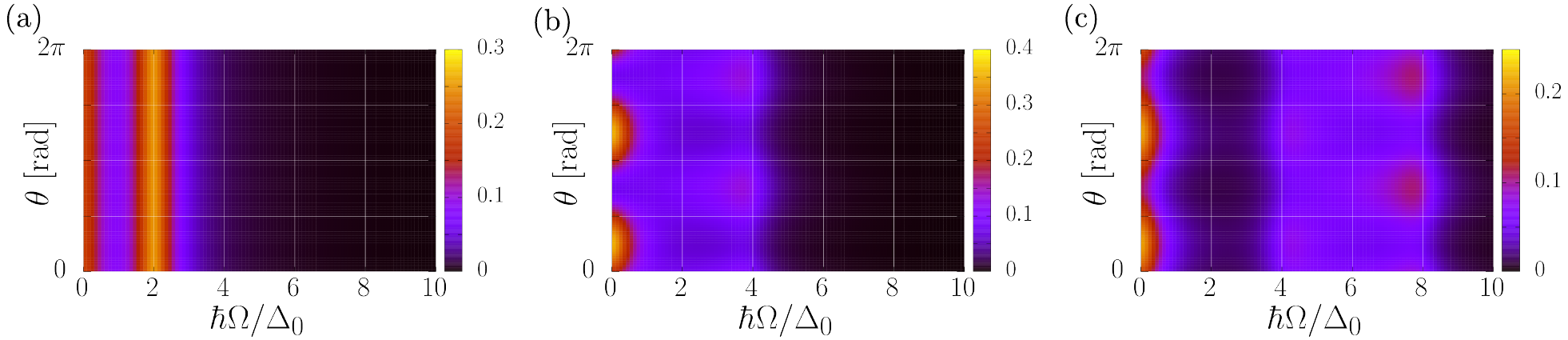}
\caption{Contour plot of enhancement of Gilbert damping in 2DEG for (a) $\alpha/\beta=0$, (b) $\alpha/\beta=1$, and (c) $\alpha/\beta=3$.
The horizontal axis is the FMR frequency $\Omega$ and the vertical axis is the azimuth angle of the spin orientation of the FI, $\theta$.}
\label{fig:contour}
\end{figure*}

Fig.~\ref{fig:increase_Gilbert_correction_term} illustrates the enhancement of the Gilbert damping, $\delta \alpha_{\rm G}$, as a function of the resonant frequency $\Omega$ for $\Gamma/\Delta_0 = 0.5$.
In particular, Fig.~\ref{fig:increase_Gilbert_correction_term} (a) and (b) shows the enhancement for $\alpha/\beta=1$ and 3, respectively.
{We note that $\delta \alpha_{\rm G}$ depends on the resonant frequency in contrast with the Gilbert damping coefficient for the bulk FI, $\alpha_{\rm G}$.}
In both cases, the enhancement $\delta \alpha_{\rm G}$ clearly depends on the spin orientation of the FI.
A peak at $\Omega = 0$ is caused by the static part $\delta \alpha_{{\rm G},1}$, while the structure at a finite $\Omega$ comes from the magnon-absorption process described by $\delta \alpha_{{\rm G},2}$.
Note that the broad structure indicated by $\delta \alpha_{{\rm G},2}$ is produced by the variation in the energy splitting $2h_{\rm eff}(\varphi)$ when the azimuth angle $\varphi$ changes from 0 to $2\pi$.
The range of $2h_{\rm eff}(\varphi)$ is obtained from Eq.~(\ref{eq:heffdef}) as $0 \le 2h_{\rm eff} \le 4\Delta_0$ for $\alpha/\beta=1$ and $4\Delta_0 \le 2 h_{\rm eff} \le 8 \Delta_0$ for $\alpha/\beta=3$.
This variation in the spin splitting corresponds to the range of the broad structure at finite $\Omega$ shown in Fig.~\ref{fig:increase_Gilbert_correction_term}~(a) and (b).

When the azimuth angle of the spin polarization in the FI, $\theta$, is varied, the enhancement of the Gilbert damping is modified, as shown in Fig.~\ref{fig:increase_Gilbert_correction_term}.
Its general properties are summarized as follows.
First, $\delta \alpha_{\rm G}$ is independent of the spin orientation of the FI for $\alpha = 0$ or $\beta = 0$.
Second, the result for $\alpha < 0$ is related to that for $\alpha > 0$ through
\begin{align}
\delta \alpha_{\rm G}(\alpha,\theta,\Omega) = \delta \alpha_{\rm G}(-\alpha,\theta-\pi/2,\Omega).
\end{align}
Third, $\delta \alpha_{\rm G}$ has symmetry relations with respect to $\theta$ as
\begin{align}
\delta \alpha_{\rm G}(\theta,\Omega) 
= \delta \alpha_{\rm G}(\theta+\pi,\Omega)
= \delta \alpha_{\rm G}(\pi/2-\theta,\Omega).
\end{align}
To see the spin-orientation dependence in more detail, we show a contour plot of $\delta \alpha_{\rm G}$ as a function of $\theta$ and $\Omega$ in Fig.~\ref{fig:contour}.
For $\alpha = 0$, $\delta \alpha_{\rm G}$ is independent of $\theta$ (Fig.~\ref{fig:contour}~(a)).
The same result is obtained for $\beta=0$ after replacing $\Delta_0$ with $k_{\rm F} \alpha$.
The spin-orientation dependence becomes strongest for $\alpha/\beta = 1$ (Fig.~\ref{fig:contour}~(b)).
For this special case, the direction of $\hat{\bm h}_{\rm eff}$ is fixed:
\begin{align}
\hat{\bm h}_{\rm eff} = \pm (1/\sqrt{2},-1/\sqrt{2},0) .
\end{align}
For the spin texture of this special case, see Fig.~\ref{fig:spin_texture}~(a).
Therefore, the $\theta$-dependent part in Eqs.~(\ref{eq:result1})-(\ref{eq:result3}) can be taken out of the integrals:
\begin{align}
\delta \alpha_{{\rm G},1} &\propto \frac{1-(\hat{\bm h}_{\rm eff}\cdot \hat{\bm m})^2}{2} , 
\\
\delta \alpha_{{\rm G},2}, \delta \alpha_{{\rm G},3} &\propto \frac{1+(\hat{\bm h}_{\rm eff}\cdot \hat{\bm m})^2}{4} ,
\end{align}
where we have used the fact that the term proportional to $\hat{\bm h}_{\rm eff}\cdot \hat{\bm m}$ vanishes after the integration with respect to $\varphi$.
From this expression, the spin-orientation dependence shown in Fig.~\ref{fig:contour}~(b) can be explained as follows.
The peak at $\Omega = 0$ that is caused by $\delta \alpha_{{\rm G},1}$ takes a maximum (a minimum) when $\hat{\bm h}_{\rm eff} \perp \hat{\bm m}$ ($\hat{\bm h}_{\rm eff} \parallel \hat{\bm m}$) or equivalently when $\theta = \pi/4, 5\pi/4$ ($\theta = 3\pi/4, 7\pi/4$).
This observation supports the conclusion that the enhancement in Gilbert damping at $\Omega = 0$ is induced by the transverse component of the effective magnetic field ${\bm h}_{\rm eff}$.
In contrast, the broad structure at finite frequencies in the range of $0 \le \hbar \Omega \le 4\Delta_0$, that is caused by $\delta \alpha_{{\rm G},2}$, takes a maximum (a minimum) when $\hat{\bm h}_{\rm eff} \parallel \hat{\bm m}=0$ ($\hat{\bm h}_{\rm eff} \perp \hat{\bm m}$).
This is consistent with the fact that this contribution comes from the magnon absorption accompanying spin-flips of the conduction electrons.
Fig.~\ref{fig:contour}~(c) shows the spin-orientation dependence for $\alpha/\beta = 3$.
Although the $\theta$ dependence cannot be expressed in a simple form for $\alpha/\beta = 3$, the qualitative features are the same as in the case of $\alpha/\beta = 1$, as indicated by comparing Fig.~\ref{fig:contour}~(b) and (c) except that the finite-frequency bread structure shifts toward the high-frequency region $4\Delta_0 \le \hbar \Omega \le 8\Delta_0$.

\subsection{Relevance to experiments}

Our results indicate that the spin-orientation dependent provides information on spin-orbit interactions in 2DEG, in which both the Rashba and Dresselhaus spin-orbit interactions coexist.
Let us estimate a necessary condition for observation of the present result.
For GaAs/AlGaAs heterostructures~\cite{Miller2003}, the magnitude of the spin-orbit interactions is given as $\alpha \sim \beta \sim 4$\ meV$\cdot$\AA, leading to $\Delta_0 = k_{\rm F}\beta \sim 0.10 \ {\rm meV}$ for the electron density $5\times 10^{11}\ {\rm cm}^{-2}$.
Because the FMR frequency for YIG under a magnetic field of 1 T is about $\hbar \Omega = 0.06 \ {\rm meV}$, the ratio $\hbar \Omega/\Delta_0$ is of order 1.
This indicates that both the elastic contribution $\delta \alpha_{{\rm G},1}$ and the magnon absorption contribution $\delta \alpha_{{\rm G},2}$ can be observed experimentally using a magnetic field of a few tesla.
Note that the Rashba spin-orbit interaction can be controlled by applying an electric field to the sample.
The amplitude of the spin-orbit interactions depends on the aspects of bulk semiconductors as well as on sample fabrication considerations.
For example, in asymmetric InAs heterostructures~\cite{Grundler2000,Sato2001}, the magnitude of the Rashba spin-orbit interaction is about $\alpha \sim 400$\ meV$\cdot$\AA, leading to $k_{\rm F}\alpha \sim 14 \ {\rm meV}$ for the electron density $10^{12}\ {\rm cm}^{-2}$.
In this case, the dependence of the spin-orientation of FI is governed by the elastic contribution $\delta \alpha_{{\rm G},1}$.
However, by using symmetric InAs heterostructures~\cite{Meier2007}, it is possible to reduce the magnitude of the Rashba spin-orbit interaction down to the same order as in GaAs/AlGaAs heterostructures.
In such heterostructures, we can also observe the contribution from magnon absorption, $\delta \alpha_{{\rm G},2}$.

\section{SUMMARY}
\label{sec:summary}
We theoretically investigated spin pumping from a ferromagnetic insulator (FI) into a two-dimensional gas (2DEG) with both Rashba and Dresselhaus spin-orbit interactions.
We considered the interfacial coupling through the tunnel Hamiltonian in which the momentum of spin excitation is conserved and derived an increase in the linewidth in a ferromagnetic resonance (FMR) experiment that is induced by the 2DEG within a second-order perturbation with respect to the interfacial coupling.
We found that there are three processes that enhance the Gilbert damping: (a) an elastic process, (b) a magnon absorption process, and (c) a magnon emission process.
The elastic process is induced by spin-flips through the transverse component of the effective magnetic field felt by conduction electrons that originate from the spin-orbit interaction in the 2DEG.
This elastic process is dominant when the FMR frequency is sufficiently low compared with the energy scale of the spin-orbit interaction.
In contrast, the magnon absorption/emission process is a dynamical one that changes the number of magnons in the FI and affects the Gilbert damping when the FMR frequency is comparable to the spin splitting energy by spin-orbit coupling in the 2DEG.
We discussed how these three processes of enhancing the Gilbert damping depend on the spin orientation in the FI.
We also showed that our results can be detected in an FMR experiment using a GaAs/AlGaAs heterostructure under a magnetic field of a few tesla.
Our work provides a helpful experimental method for the detection of spin texture of conduction electrons at the Fermi surface.

\section*{ACKNOWLEDGMENTS}
We would like to thank Dr. Y. Ominato for fruitful discussions.
T. K. acknowledges support from the Japan Society for the Promotion of Science (JSPS KAKENHI Grant No.~JP20K03831). M. M. is financially supported by a Grant-in-Aid for Scientific Research (Grants No.~JP20H01863 and No.~JP21H04565) from MEXT, Japan. 

\appendix

\section{IMPURITY SCATTERING}
\label{sec:ImpurityScattering}

\begin{figure}[tb]
\centering
\includegraphics[width=40mm]{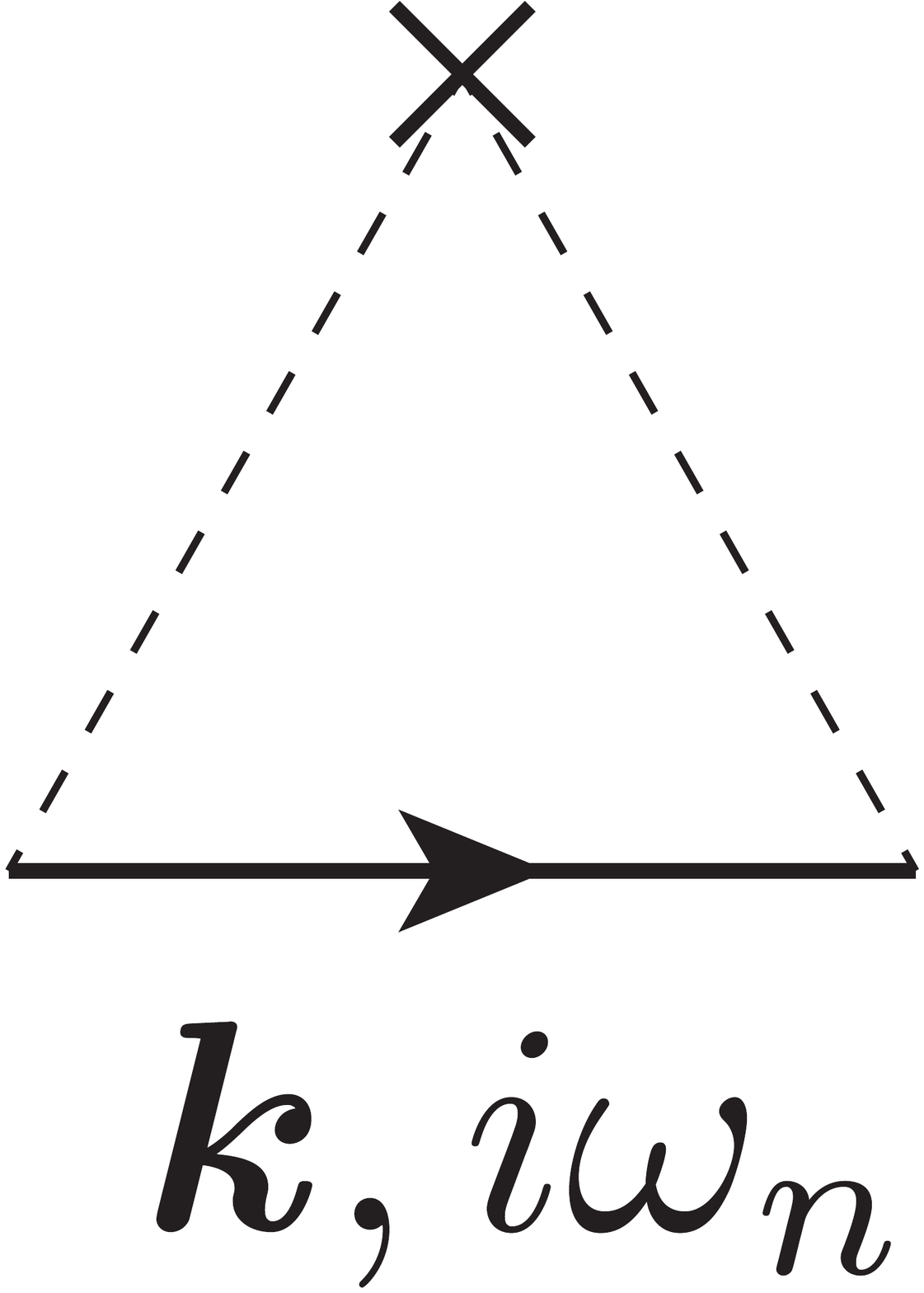}
\caption{Feynman diagram of second-order perturbation with respect to the impurity potential.}
\label{fig:diagram}
\end{figure}

In this study, we consider the effect of impurity scattering within a second-order perturbation with respect to an impurity potential by taking a random average.
This approximation corresponds to the Born approximation, whose diagram is shown in Fig.~\ref{fig:diagram}.
In this approximation, the temperature Green's function is written as
\begin{align}
(\hat{g}({\bm k},i\omega_n))^{-1} &= 
(\hat{g}_0({\bm k},i\omega_n))^{-1}-\hat{\Gamma}(i\omega_n), 
\label{eq:Dyson} \\
\hat{\Gamma}(i\omega_n) &= n_i u^2 \int \frac{d^2{\bm k}}{(2\pi)^2} \, \hat{g}_0({\bm k},i\omega_n),
\label{eq:Gamma}
\end{align}
where $n_i$ is the number of impurity sites.
We assume that the scattering rate is much smaller than the bandwidth of the conduction electrons.
Accordingly, the self-energy is calculated as
\begin{align}
\hat{\Gamma}(i\omega_n) &= -i\frac{n_i u^2 k_{\rm F}}{2v_{\rm F}} {\rm sgn}(\omega_n) \hat{I} 
\equiv -i \frac{\Gamma}{2} {\rm sgn}(\omega_n) \hat{I},
\end{align}
where $\Gamma$ denotes the impurity scattering rate.
Using the Dyson equation (\ref{eq:Dyson}), the retarded component of the Green's function is obtained as Eq.~(\ref{eq:gkwimp}).

\section{SPIN-WAVE APPROXIMATION}
\label{app:SpinWave}

We derive the Hamiltonian within the spin-wave approximation by using the Holstein-Primakov transformation.
For $S_{0}\gg 1$, it is written as
\begin{align}
    S_i^{x'-} &\simeq \sqrt{2S_0} b_i^\dagger , \\
    S_i^{x'+} &\simeq \sqrt{2S_0} b_i, \\
    S_i^{x'} &= S_0-b_i^\dagger b_i ,
\end{align}
where $b_i$ ($b_i^\dagger$) is an annihilation (creation) operator defined at site $i$.
We replace the spin operators with these boson operators and take the Fourier transform,
\begin{align}
    b_i = \frac{1}{\sqrt{N_{\rm F}}} \sum_{{\bm k}} e^{i{\bm k}\cdot {\bm r}_i} b_{\bm k},
\end{align}
where $N_{\rm F}$ is the number of unit cells in the FI.
The Hamiltonian of the FI is modified into Eqs.~(\ref{eq:HFI1}) and (\ref{eq:HFI2}).
When we consider the cubic lattice model with only the nearest-neighbor exchange coupling $J$, the dispersion is given as $\hbar \omega_{\bm k} = \hbar \omega_{\bm k}^{(0)} + \hbar \gamma h_{\rm dc}$, where
\begin{align}
\hbar \omega_{\bm k}^{(0)}
&= 2JS_0(3-\cos(k_x a)-\cos(k_y a)-\cos(k_z a))  \nonumber \\
&\simeq JS_0a^2 {\bm k}^2 .
\end{align}
and $a$ is a lattice constant of the FI. The last equation is the long-wavelength approximation.

\section{DERIVATION OF EQUATION~(\ref{eq:sigmaresult1})}
\label{app:CalcSpin}

Here, we derive Eq.~(\ref{eq:sigmaresult1}).
We rewrite Green's function of the conduction electrons as
\begin{align}
\hat{g}(\bm k,i\omega_{n}) &= \frac{1}{D(i\omega_{n})} \left[ A(i\omega_{n})\hat{I}+{\bm b}\cdot {\bm \sigma}\right], 
\end{align}
where ${\bm a}=(-\sin\theta,\cos\theta,i)$ and ${\bm b}=-{\bm h}_{\rm eff}$.
Then, the trace in Eq.~(\ref{eq:self_Sigma}) is rewritten as
\begin{align}
&I \equiv {\rm Tr} \left[ \hat{\sigma}^{x'-}
\hat{g}({\bm k},i\omega_m)\hat{\sigma}^{x'+}
\hat{g}({\bm k},i\omega_m+i\omega_n) \right]\nonumber\\
&=\frac{1}{DD'}
{\rm Tr}\Bigl[{\bm a}^*\cdot{\bm \sigma} (A\hat{I}+{\bm b}\cdot{\bm \sigma}){\bm a}\cdot {\bm \sigma} (A'\hat{I}+{\bm b}\cdot {\bm \sigma}) \Bigl].
\end{align}
Using the identity,
\begin{align}
& ({\bm a}\cdot {\bm \sigma})({\bm b}\cdot {\bm \sigma}) = ({\bm a}\cdot {\bm b}) \hat{I} + i ({\bm a}\times {\bm b})\cdot{\bm \sigma}, \\
& {\rm Tr} \, [\sigma_a] = 0, \quad \quad {\rm Tr} \, [\hat{I}] = 2,
\label{eq:PauliFormula}
\end{align}
a straightforward calculation gives
\begin{align}
I &=\frac{2}{DD'} \Bigl[
A A' {\bm a}^* \cdot {\bm a}
+ iA' ({\bm a}^*\times {\bm b})\cdot {\bm a}
+ iA {\bm a}^*\cdot ({\bm a}\times {\bm b}) \nonumber \\
& \hspace{10mm} - ({\bm a}^*\times {\bm b})\cdot({\bm a}\times {\bm b})
+ ({\bm a}^*\cdot {\bm b})
({\bm a}\cdot {\bm b}) \Bigr] .
\end{align}
We obtain Eq.~(\ref{eq:sigmaresult1}) by substituting the explicit forms of ${\bm a}$ and ${\bm b}$.

\section{ANALYTIC CONTINUATION}
\label{app:AnalyticContinuation}
Here, we perform the summation in the self-energy by using analytic continuation.
Using the identities,
\begin{align}
& \frac{A}{D} = \frac{1}{2} \sum_{\nu=\pm} \frac{1}{i\hbar \omega_m-E_{\bm k}^\nu+i\Gamma/2\,{\rm sgn}(\omega_m)}, \\
& \frac{{\bm h}_{\rm eff}\cdot \hat{\bm m}}{D} = \frac{1}{2}
\sum_{\nu = \pm} 
\frac{\nu \hat{\bm h}_{\rm eff}\cdot \hat{\bm m}}{i\hbar \omega_m-E_{\bm k}^\nu + i\Gamma/2\,{\rm sgn}(\omega_m)},
\end{align}
and the counterparts for $A'$ and $D'$, the self-energy is rewritten as
\begin{align}
& \Sigma({\bm 0},i\omega_n) = \frac{|T_{\bm 0}|^2}{4}
\sum_{{\bm k}}\sum_{\nu=\pm} \sum_{\nu'=\pm} {(1-\nu \hat{\bm h}_{\rm eff}\cdot \hat{\bm m})}\nonumber \\
&\hspace{40mm}{\times(1+\nu' \hat{\bm h}_{\rm eff}\cdot \hat{\bm m})} I_{{\bm k}\nu\nu'}, \\
& I_{{\bm k}\nu\nu'} = \frac{1}{\tempinv} \sum_{i\omega_m}
\frac{{1}}{i\hbar \omega_m-E_{\bm k}^\nu+i\Gamma/2 \,{\rm sgn}(\omega_m)} \nonumber \\
& \hspace{5mm} \times 
\frac{{1}}{i\hbar \omega_m+i\hbar \omega_n-E_{\bm k}^{\nu'}+i\Gamma/2 \, {\rm sgn}(\omega_m+\omega_n)},
\end{align}
where $\hat{\bm h}_{\rm eff} = {\bm h}_{\rm eff}/h_{\rm eff}$.
By using analytic continuation, $I_{{\bm k}\nu \nu'}$ can be expressed as a contour integral,
\begin{align}
I_{{\bm k}\nu\nu'} &= - \int_{C} \frac{dz}{2\pi i} f(z)
\frac{1}{z-E_{\bm k}^\nu + i\Gamma/2 \, {\rm sgn}({\rm Im}\, z)} \nonumber \\
& \times \frac{1}{z+i\hbar \omega_n-E_{\bm k}^{\nu'} +i\Gamma/2 \, {\rm sgn}({\rm Im}\,z+\omega_n)},
\end{align}
where $f(z) = (e^{\tempinv z}+1)^{-1}$ and $C$ is a contour surrounding the poles of $f(z)$.

\begin{figure}[t]
\centering
\includegraphics[width=85mm]{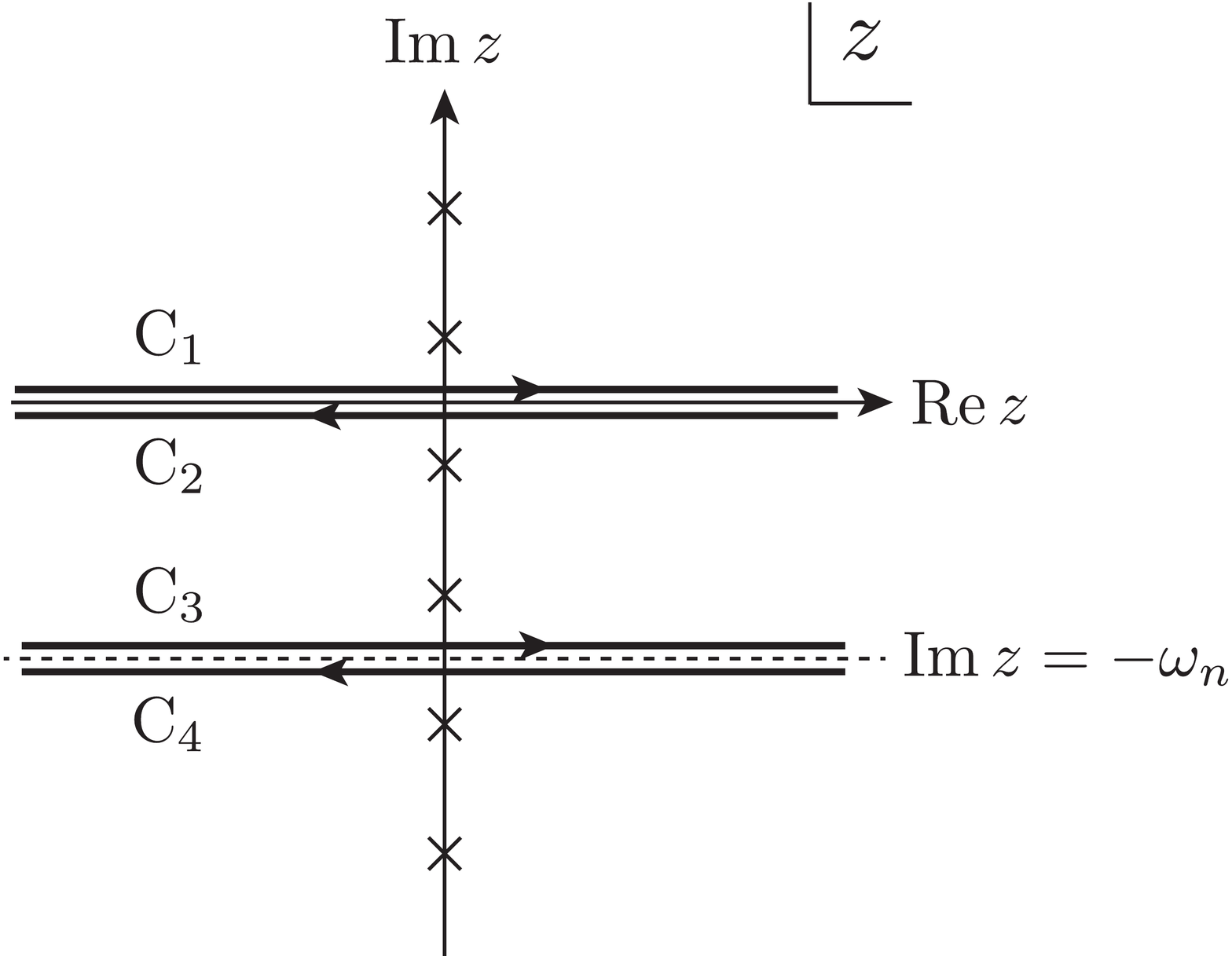}
\caption{Contour on the complex plane.}
\label{fig:ComplexContour}
\end{figure}

We modify the contour $C$ to be a sum of $C_1$, $C_2$, $C_3$, and $C_4$, as shown in Fig.~\ref{fig:ComplexContour}, and change the integration variable to $z=E+i\eta$ for $C_1$, $z=E - i\eta$ for $C_2$, $z=E -i\omega_n + i\eta$ for $C_3$, and $z= E -i\omega_n - i\eta$ for $C_4$, where $\eta$ is a positive infinitesimal.
Then, we obtain
\begin{align}
& I_{{\bm k}\nu\nu'} = - \int \frac{dE}{2\pi i} f(E) \nonumber \\
& \times \Biggl[ \frac{-i\Gamma}{(E -E_{\bm k}^{\nu})^2+(\Gamma/2)^2}\times 
\frac{1}{E+i\hbar \omega_n-E_{\bm k}^{\nu'}+i\Gamma/2}  
\nonumber \\
& \hspace{5mm} + 
\frac{1}{E-i\hbar \omega_n-E_{\bm k}^{\nu}-i\Gamma/2} \times 
\frac{-i\Gamma}{(E-E_{\bm k}^{\nu'})^2+(\Gamma/2)^2}
 \Biggr] .
\end{align}
By changing the variable to $E'=E-E_{\bm k}^{\nu}$ for the first term and to $E'=-(E-E_{\bm k}^{\nu'})$ for the second term, we obtain
\begin{align}
I_{{\bm k} \nu\nu'} &= - \int \frac{dE'}{2\pi i} 
\frac{-i\Gamma}{E'^2+(\Gamma/2)^2}  \nonumber \\
& \hspace{5mm} \times 
\Biggl[ \frac{f(E'+E_{\bm k}^{\nu})-
f(-E'+E_{\bm k}^{\nu'})}{E'+i\hbar \omega_n+E_{\bm k}^{\nu}-E_{\bm k}^{\nu'}+i\Gamma/2} \Biggr] .
\end{align}
The summation with respect to the wavenumber can be replaced with an integral,
\begin{align}
\frac{1}{{\cal A}} \sum_{\bm k} I_{{\bm k}\nu\nu'}
\simeq D(\epsilon_{\rm F}) \int_{-\infty}^\infty d\xi \int_0^{2\pi}  \frac{d\varphi}{2\pi}
I_{{\bm k}\nu \nu'},
\end{align}
where ${\cal A}$ is the junction area and $\xi\equiv \xi_{\bm k}$.
Using the integral formulas,
\begin{align}
& \int_{-\infty}^\infty d\xi \, (f(E'+E^{\nu}_{\bm k})-f(-E'+E^{\nu'}_{\bm k})) 
\nonumber \\
&\hspace{10mm} = -(2E'+E^{\nu}_{\bm k}-E^{\nu'}_{\bm k}), \\
& \int_{-\infty}^{\infty} \frac{dx}{2\pi}
\frac{a^2}{x^2+(a/2)^2}
\frac{x+b/2}{(x+b+c)^2+(a/2)^2} \nonumber \\
& \hspace{10mm} = -
\frac{ac}{(b+c)^2+a^2}, \quad (a>0),
\end{align}
we finally obtain
\begin{align}
& {\rm Im} \, \Sigma^R({\bm 0},\omega) \nonumber \\
&= -\frac{|T_{\bm 0}|^2 {\cal A} D(\epsilon_{\rm F})}{4} 
\sum_{\nu,\nu'} 
\int_0^{2\pi} \frac{d\varphi}{2\pi}
(1-\nu \hat{\bm h}_{\rm eff}(\varphi) \cdot \hat{\bm m}) \nonumber \\
& \hspace{5mm} \times (1+\nu' \hat{\bm h}_{\rm eff}(\varphi) \cdot \hat{\bm m})
\frac{\Gamma \hbar \omega}{(\hbar \omega + E_{\bm k}^{\nu}-E_{\bm k}^{\nu'})^2+\Gamma^2}.
\end{align}
Note that the final result does not depend on the temperature.
This feature emerges when the density of states for conduction electrons is approximated as being constant near the Fermi energy.
In general, one can derive a small temperature-dependent correction by using a Sommerfeld expansion that takes into account the energy dependence of the density of states.

\bibliography{ref}

\begin{thebibliography}{50}%
\makeatletter
\providecommand \@ifxundefined [1]{%
 \@ifx{#1\undefined}
}%
\providecommand \@ifnum [1]{%
 \ifnum #1\expandafter \@firstoftwo
 \else \expandafter \@secondoftwo
 \fi
}%
\providecommand \@ifx [1]{%
 \ifx #1\expandafter \@firstoftwo
 \else \expandafter \@secondoftwo
 \fi
}%
\providecommand \natexlab [1]{#1}%
\providecommand \enquote  [1]{``#1''}%
\providecommand \bibnamefont  [1]{#1}%
\providecommand \bibfnamefont [1]{#1}%
\providecommand \citenamefont [1]{#1}%
\providecommand \href@noop [0]{\@secondoftwo}%
\providecommand \href [0]{\begingroup \@sanitize@url \@href}%
\providecommand \@href[1]{\@@startlink{#1}\@@href}%
\providecommand \@@href[1]{\endgroup#1\@@endlink}%
\providecommand \@sanitize@url [0]{\catcode `\\12\catcode `\$12\catcode
  `\&12\catcode `\#12\catcode `\^12\catcode `\_12\catcode `\%12\relax}%
\providecommand \@@startlink[1]{}%
\providecommand \@@endlink[0]{}%
\providecommand \url  [0]{\begingroup\@sanitize@url \@url }%
\providecommand \@url [1]{\endgroup\@href {#1}{\urlprefix }}%
\providecommand \urlprefix  [0]{URL }%
\providecommand \Eprint [0]{\href }%
\providecommand \doibase [0]{http://dx.doi.org/}%
\providecommand \selectlanguage [0]{\@gobble}%
\providecommand \bibinfo  [0]{\@secondoftwo}%
\providecommand \bibfield  [0]{\@secondoftwo}%
\providecommand \translation [1]{[#1]}%
\providecommand \BibitemOpen [0]{}%
\providecommand \bibitemStop [0]{}%
\providecommand \bibitemNoStop [0]{.\EOS\space}%
\providecommand \EOS [0]{\spacefactor3000\relax}%
\providecommand \BibitemShut  [1]{\csname bibitem#1\endcsname}%
\let\auto@bib@innerbib\@empty
\bibitem [{\citenamefont {Tserkovnyak}\ \emph {et~al.}(2002)\citenamefont
  {Tserkovnyak}, \citenamefont {Brataas},\ and\ \citenamefont
  {Bauer}}]{Tserkovnyak2002}%
  \BibitemOpen
  \bibfield  {author} {\bibinfo {author} {\bibfnamefont {Y.}~\bibnamefont
  {Tserkovnyak}}, \bibinfo {author} {\bibfnamefont {A.}~\bibnamefont
  {Brataas}}, \ and\ \bibinfo {author} {\bibfnamefont {G.~E.~W.}\ \bibnamefont
  {Bauer}},\ }\href {\doibase 10.1103/PhysRevLett.88.117601} {\bibfield
  {journal} {\bibinfo  {journal} {Phys. Rev. Lett.}\ }\textbf {\bibinfo
  {volume} {88}},\ \bibinfo {pages} {117601} (\bibinfo {year}
  {2002})}\BibitemShut {NoStop}%
\bibitem [{\citenamefont {Tserkovnyak}\ \emph {et~al.}(2005)\citenamefont
  {Tserkovnyak}, \citenamefont {Brataas}, \citenamefont {Bauer},\ and\
  \citenamefont {Halperin}}]{Tserkovnyak2005}%
  \BibitemOpen
  \bibfield  {author} {\bibinfo {author} {\bibfnamefont {Y.}~\bibnamefont
  {Tserkovnyak}}, \bibinfo {author} {\bibfnamefont {A.}~\bibnamefont
  {Brataas}}, \bibinfo {author} {\bibfnamefont {G.~E.~W.}\ \bibnamefont
  {Bauer}}, \ and\ \bibinfo {author} {\bibfnamefont {B.~I.}\ \bibnamefont
  {Halperin}},\ }\href {\doibase 10.1103/RevModPhys.77.1375} {\bibfield
  {journal} {\bibinfo  {journal} {Rev. Mod. Phys.}\ }\textbf {\bibinfo {volume}
  {77}},\ \bibinfo {pages} {1375} (\bibinfo {year} {2005})}\BibitemShut
  {NoStop}%
\bibitem [{\citenamefont {Mizukami}\ \emph {et~al.}(2001)\citenamefont
  {Mizukami}, \citenamefont {Ando},\ and\ \citenamefont
  {Miyazaki}}]{Mizukami2001}%
  \BibitemOpen
  \bibfield  {author} {\bibinfo {author} {\bibfnamefont {S.}~\bibnamefont
  {Mizukami}}, \bibinfo {author} {\bibfnamefont {Y.}~\bibnamefont {Ando}}, \
  and\ \bibinfo {author} {\bibfnamefont {T.}~\bibnamefont {Miyazaki}},\ }\href
  {\doibase https://doi.org/10.1143/JJAP.40.580} {\bibfield  {journal}
  {\bibinfo  {journal} {Jpn. J. Appl. Phys.}\ }\textbf {\bibinfo {volume}
  {40}},\ \bibinfo {pages} {580} (\bibinfo {year} {2001})}\BibitemShut
  {NoStop}%
\bibitem [{\citenamefont {Mizukami}\ \emph {et~al.}(2002)\citenamefont
  {Mizukami}, \citenamefont {Ando},\ and\ \citenamefont
  {Miyazaki}}]{Mizukami2002}%
  \BibitemOpen
  \bibfield  {author} {\bibinfo {author} {\bibfnamefont {S.}~\bibnamefont
  {Mizukami}}, \bibinfo {author} {\bibfnamefont {Y.}~\bibnamefont {Ando}}, \
  and\ \bibinfo {author} {\bibfnamefont {T.}~\bibnamefont {Miyazaki}},\ }\href
  {\doibase 10.1103/PhysRevB.66.104413} {\bibfield  {journal} {\bibinfo
  {journal} {Phys. Rev. B}\ }\textbf {\bibinfo {volume} {66}},\ \bibinfo
  {pages} {104413} (\bibinfo {year} {2002})}\BibitemShut {NoStop}%
\bibitem [{\citenamefont {Saitoh}\ \emph {et~al.}(2006)\citenamefont {Saitoh},
  \citenamefont {Ueda}, \citenamefont {Miyajima},\ and\ \citenamefont
  {Tatara}}]{Saitoh2006}%
  \BibitemOpen
  \bibfield  {author} {\bibinfo {author} {\bibfnamefont {E.}~\bibnamefont
  {Saitoh}}, \bibinfo {author} {\bibfnamefont {M.}~\bibnamefont {Ueda}},
  \bibinfo {author} {\bibfnamefont {H.}~\bibnamefont {Miyajima}}, \ and\
  \bibinfo {author} {\bibfnamefont {G.}~\bibnamefont {Tatara}},\ }\href
  {\doibase 10.1063/1.2199473} {\bibfield  {journal} {\bibinfo  {journal}
  {Appl. Phys. Lett.}\ }\textbf {\bibinfo {volume} {88}},\ \bibinfo {pages}
  {182509} (\bibinfo {year} {2006})}\BibitemShut {NoStop}%
\bibitem [{\citenamefont {Ando}\ \emph {et~al.}(2008)\citenamefont {Ando},
  \citenamefont {Kajiwara}, \citenamefont {Takahashi}, \citenamefont {Maekawa},
  \citenamefont {Takemoto}, \citenamefont {Takatsu},\ and\ \citenamefont
  {Saitoh}}]{Ando2008}%
  \BibitemOpen
  \bibfield  {author} {\bibinfo {author} {\bibfnamefont {K.}~\bibnamefont
  {Ando}}, \bibinfo {author} {\bibfnamefont {Y.}~\bibnamefont {Kajiwara}},
  \bibinfo {author} {\bibfnamefont {S.}~\bibnamefont {Takahashi}}, \bibinfo
  {author} {\bibfnamefont {S.}~\bibnamefont {Maekawa}}, \bibinfo {author}
  {\bibfnamefont {K.}~\bibnamefont {Takemoto}}, \bibinfo {author}
  {\bibfnamefont {M.}~\bibnamefont {Takatsu}}, \ and\ \bibinfo {author}
  {\bibfnamefont {E.}~\bibnamefont {Saitoh}},\ }\href {\doibase
  10.1103/PhysRevB.78.014413} {\bibfield  {journal} {\bibinfo  {journal} {Phys.
  Rev. B}\ }\textbf {\bibinfo {volume} {78}},\ \bibinfo {pages} {014413}
  (\bibinfo {year} {2008})}\BibitemShut {NoStop}%
\bibitem [{\citenamefont {Kajiwara}\ \emph {et~al.}(2010)\citenamefont
  {Kajiwara}, \citenamefont {Harii}, \citenamefont {Takahashi}, \citenamefont
  {Ohe}, \citenamefont {Uchida}, \citenamefont {Mizuguchi}, \citenamefont
  {Umezawa}, \citenamefont {Kawai}, \citenamefont {Ando}, \citenamefont
  {Takanashi}, \citenamefont {Maekawa},\ and\ \citenamefont
  {Saitoh}}]{Kajiwara2010}%
  \BibitemOpen
  \bibfield  {author} {\bibinfo {author} {\bibfnamefont {Y.}~\bibnamefont
  {Kajiwara}}, \bibinfo {author} {\bibfnamefont {K.}~\bibnamefont {Harii}},
  \bibinfo {author} {\bibfnamefont {S.}~\bibnamefont {Takahashi}}, \bibinfo
  {author} {\bibfnamefont {J.}~\bibnamefont {Ohe}}, \bibinfo {author}
  {\bibfnamefont {K.}~\bibnamefont {Uchida}}, \bibinfo {author} {\bibfnamefont
  {M.}~\bibnamefont {Mizuguchi}}, \bibinfo {author} {\bibfnamefont
  {H.}~\bibnamefont {Umezawa}}, \bibinfo {author} {\bibfnamefont
  {H.}~\bibnamefont {Kawai}}, \bibinfo {author} {\bibfnamefont
  {K.}~\bibnamefont {Ando}}, \bibinfo {author} {\bibfnamefont {K.}~\bibnamefont
  {Takanashi}}, \bibinfo {author} {\bibfnamefont {S.}~\bibnamefont {Maekawa}},
  \ and\ \bibinfo {author} {\bibfnamefont {E.}~\bibnamefont {Saitoh}},\ }\href
  {\doibase 10.1038/nature08876} {\bibfield  {journal} {\bibinfo  {journal}
  {Nat.}\ }\textbf {\bibinfo {volume} {464}},\ \bibinfo {pages} {262} (\bibinfo
  {year} {2010})}\BibitemShut {NoStop}%
\bibitem [{\citenamefont {\ifmmode \check{Z}\else
  \v{Z}\fi{}uti\ifmmode~\acute{c}\else \'{c}\fi{}}\ \emph
  {et~al.}(2004)\citenamefont {\ifmmode \check{Z}\else
  \v{Z}\fi{}uti\ifmmode~\acute{c}\else \'{c}\fi{}}, \citenamefont {Fabian},\
  and\ \citenamefont {Das~Sarma}}]{Zutic2004}%
  \BibitemOpen
  \bibfield  {author} {\bibinfo {author} {\bibfnamefont {I.}~\bibnamefont
  {\ifmmode \check{Z}\else \v{Z}\fi{}uti\ifmmode~\acute{c}\else \'{c}\fi{}}},
  \bibinfo {author} {\bibfnamefont {J.}~\bibnamefont {Fabian}}, \ and\ \bibinfo
  {author} {\bibfnamefont {S.}~\bibnamefont {Das~Sarma}},\ }\href {\doibase
  10.1103/RevModPhys.76.323} {\bibfield  {journal} {\bibinfo  {journal} {Rev.
  Mod. Phys.}\ }\textbf {\bibinfo {volume} {76}},\ \bibinfo {pages} {323}
  (\bibinfo {year} {2004})}\BibitemShut {NoStop}%
\bibitem [{\citenamefont {Awschalom}\ and\ \citenamefont
  {Flatt{\'e}}(2007)}]{Awschalom2007}%
  \BibitemOpen
  \bibfield  {author} {\bibinfo {author} {\bibfnamefont {D.~D.}\ \bibnamefont
  {Awschalom}}\ and\ \bibinfo {author} {\bibfnamefont {M.~E.}\ \bibnamefont
  {Flatt{\'e}}},\ }\href {\doibase https://doi.org/10.1038/nphys551} {\bibfield
   {journal} {\bibinfo  {journal} {Nat. Phys.}\ }\textbf {\bibinfo {volume}
  {3}},\ \bibinfo {pages} {153} (\bibinfo {year} {2007})}\BibitemShut {NoStop}%
\bibitem [{\citenamefont {Datta}\ and\ \citenamefont {Das}(1990)}]{Datta1990}%
  \BibitemOpen
  \bibfield  {author} {\bibinfo {author} {\bibfnamefont {S.}~\bibnamefont
  {Datta}}\ and\ \bibinfo {author} {\bibfnamefont {B.}~\bibnamefont {Das}},\
  }\href {\doibase 10.1063/1.102730} {\bibfield  {journal} {\bibinfo  {journal}
  {Appl. Phys. Lett.}\ }\textbf {\bibinfo {volume} {56}},\ \bibinfo {pages}
  {665} (\bibinfo {year} {1990})}\BibitemShut {NoStop}%
\bibitem [{\citenamefont {Srisongmuang}\ \emph {et~al.}(2008)\citenamefont
  {Srisongmuang}, \citenamefont {Pairor},\ and\ \citenamefont
  {Berciu}}]{Srisongmuang2008}%
  \BibitemOpen
  \bibfield  {author} {\bibinfo {author} {\bibfnamefont {B.}~\bibnamefont
  {Srisongmuang}}, \bibinfo {author} {\bibfnamefont {P.}~\bibnamefont
  {Pairor}}, \ and\ \bibinfo {author} {\bibfnamefont {M.}~\bibnamefont
  {Berciu}},\ }\href {\doibase 10.1103/PhysRevB.78.155317} {\bibfield
  {journal} {\bibinfo  {journal} {Phys. Rev. B}\ }\textbf {\bibinfo {volume}
  {78}},\ \bibinfo {pages} {155317} (\bibinfo {year} {2008})}\BibitemShut
  {NoStop}%
\bibitem [{\citenamefont {Akabori}\ \emph {et~al.}(2012)\citenamefont
  {Akabori}, \citenamefont {Hidaka}, \citenamefont {Iwase}, \citenamefont
  {Yamada},\ and\ \citenamefont {Ekenberg}}]{Akabori2012}%
  \BibitemOpen
  \bibfield  {author} {\bibinfo {author} {\bibfnamefont {M.}~\bibnamefont
  {Akabori}}, \bibinfo {author} {\bibfnamefont {S.}~\bibnamefont {Hidaka}},
  \bibinfo {author} {\bibfnamefont {H.}~\bibnamefont {Iwase}}, \bibinfo
  {author} {\bibfnamefont {S.}~\bibnamefont {Yamada}}, \ and\ \bibinfo {author}
  {\bibfnamefont {U.}~\bibnamefont {Ekenberg}},\ }\href {\doibase
  https://doi.org/10.1063/1.4766749} {\bibfield  {journal} {\bibinfo  {journal}
  {J. Appl. Phys.}\ }\textbf {\bibinfo {volume} {112}},\ \bibinfo {pages}
  {113711} (\bibinfo {year} {2012})}\BibitemShut {NoStop}%
\bibitem [{\citenamefont {Feng}\ \emph {et~al.}(2017)\citenamefont {Feng},
  \citenamefont {Shen}, \citenamefont {Yang}, \citenamefont {Wang},
  \citenamefont {Zeng}, \citenamefont {Wu}, \citenamefont {Chintalapati},\ and\
  \citenamefont {Chang}}]{Feng2017}%
  \BibitemOpen
  \bibfield  {author} {\bibinfo {author} {\bibfnamefont {Y.~P.}\ \bibnamefont
  {Feng}}, \bibinfo {author} {\bibfnamefont {L.}~\bibnamefont {Shen}}, \bibinfo
  {author} {\bibfnamefont {M.}~\bibnamefont {Yang}}, \bibinfo {author}
  {\bibfnamefont {A.}~\bibnamefont {Wang}}, \bibinfo {author} {\bibfnamefont
  {M.}~\bibnamefont {Zeng}}, \bibinfo {author} {\bibfnamefont {Q.}~\bibnamefont
  {Wu}}, \bibinfo {author} {\bibfnamefont {S.}~\bibnamefont {Chintalapati}}, \
  and\ \bibinfo {author} {\bibfnamefont {C.-R.}\ \bibnamefont {Chang}},\ }\href
  {\doibase https://doi.org/10.1002/wcms.1313} {\bibfield  {journal} {\bibinfo
  {journal} {WIREs Comput. Mol. Sci.}\ }\textbf {\bibinfo {volume} {7}},\
  \bibinfo {pages} {e1313} (\bibinfo {year} {2017})}\BibitemShut {NoStop}%
\bibitem [{\citenamefont {Bychkov}\ and\ \citenamefont
  {Rashba}(1984)}]{Bychkov1984}%
  \BibitemOpen
  \bibfield  {author} {\bibinfo {author} {\bibfnamefont {Y.~A.}\ \bibnamefont
  {Bychkov}}\ and\ \bibinfo {author} {\bibfnamefont {E.~I.}\ \bibnamefont
  {Rashba}},\ }\href {\doibase 10.1088/0022-3719/17/33/015} {\bibfield
  {journal} {\bibinfo  {journal} {J. Phys. C: Solid State Phys.}\ }\textbf
  {\bibinfo {volume} {17}},\ \bibinfo {pages} {6039} (\bibinfo {year}
  {1984})}\BibitemShut {NoStop}%
\bibitem [{\citenamefont {Rashba}(2015)}]{Rashba2015}%
  \BibitemOpen
  \bibfield  {author} {\bibinfo {author} {\bibfnamefont {E.~I.}\ \bibnamefont
  {Rashba}},\ }\href {\doibase 10.1016/j.elspec.2014.10.002} {\bibfield
  {journal} {\bibinfo  {journal} {J. Electron Spectros. Relat. Phenomena}\
  }\textbf {\bibinfo {volume} {201}},\ \bibinfo {pages} {4} (\bibinfo {year}
  {2015})}\BibitemShut {NoStop}%
\bibitem [{\citenamefont {Dresselhaus}(1955)}]{Dresselhaus1955}%
  \BibitemOpen
  \bibfield  {author} {\bibinfo {author} {\bibfnamefont {G.}~\bibnamefont
  {Dresselhaus}},\ }\href {\doibase 10.1103/PhysRev.100.580} {\bibfield
  {journal} {\bibinfo  {journal} {Phys. Rev.}\ }\textbf {\bibinfo {volume}
  {100}},\ \bibinfo {pages} {580} (\bibinfo {year} {1955})}\BibitemShut
  {NoStop}%
\bibitem [{\citenamefont {La~Rocca}\ \emph {et~al.}(1988)\citenamefont
  {La~Rocca}, \citenamefont {Kim},\ and\ \citenamefont
  {Rodriguez}}]{Rocca1988}%
  \BibitemOpen
  \bibfield  {author} {\bibinfo {author} {\bibfnamefont {G.~C.}\ \bibnamefont
  {La~Rocca}}, \bibinfo {author} {\bibfnamefont {N.}~\bibnamefont {Kim}}, \
  and\ \bibinfo {author} {\bibfnamefont {S.}~\bibnamefont {Rodriguez}},\ }\href
  {\doibase 10.1103/PhysRevB.38.7595} {\bibfield  {journal} {\bibinfo
  {journal} {Phys. Rev. B}\ }\textbf {\bibinfo {volume} {38}},\ \bibinfo
  {pages} {7595} (\bibinfo {year} {1988})}\BibitemShut {NoStop}%
\bibitem [{\citenamefont {Manchon}\ \emph {et~al.}(2015)\citenamefont
  {Manchon}, \citenamefont {Koo}, \citenamefont {Nitta}, \citenamefont
  {Frolov},\ and\ \citenamefont {Duine}}]{Manchon2015}%
  \BibitemOpen
  \bibfield  {author} {\bibinfo {author} {\bibfnamefont {A.}~\bibnamefont
  {Manchon}}, \bibinfo {author} {\bibfnamefont {H.~C.}\ \bibnamefont {Koo}},
  \bibinfo {author} {\bibfnamefont {J.}~\bibnamefont {Nitta}}, \bibinfo
  {author} {\bibfnamefont {S.~M.}\ \bibnamefont {Frolov}}, \ and\ \bibinfo
  {author} {\bibfnamefont {R.~A.}\ \bibnamefont {Duine}},\ }\href {\doibase
  https://doi.org/10.1038/nmat4360} {\bibfield  {journal} {\bibinfo  {journal}
  {Nat. Mater.}\ }\textbf {\bibinfo {volume} {14}},\ \bibinfo {pages} {871}
  (\bibinfo {year} {2015})}\BibitemShut {NoStop}%
\bibitem [{\citenamefont {Nitta}\ \emph {et~al.}(1999)\citenamefont {Nitta},
  \citenamefont {Meijer},\ and\ \citenamefont {Takayanagi}}]{Nitta1999}%
  \BibitemOpen
  \bibfield  {author} {\bibinfo {author} {\bibfnamefont {J.}~\bibnamefont
  {Nitta}}, \bibinfo {author} {\bibfnamefont {F.~E.}\ \bibnamefont {Meijer}}, \
  and\ \bibinfo {author} {\bibfnamefont {H.}~\bibnamefont {Takayanagi}},\
  }\href {\doibase https://doi.org/10.1063/1.124485} {\bibfield  {journal}
  {\bibinfo  {journal} {Appl. Phys. Lett.}\ }\textbf {\bibinfo {volume} {75}},\
  \bibinfo {pages} {695} (\bibinfo {year} {1999})}\BibitemShut {NoStop}%
\bibitem [{\citenamefont {Ionicioiu}\ and\ \citenamefont
  {D'Amico}(2003)}]{Ionicioiu2003}%
  \BibitemOpen
  \bibfield  {author} {\bibinfo {author} {\bibfnamefont {R.}~\bibnamefont
  {Ionicioiu}}\ and\ \bibinfo {author} {\bibfnamefont {I.}~\bibnamefont
  {D'Amico}},\ }\href {\doibase 10.1103/PhysRevB.67.041307} {\bibfield
  {journal} {\bibinfo  {journal} {Phys. Rev. B}\ }\textbf {\bibinfo {volume}
  {67}},\ \bibinfo {pages} {041307(R)} (\bibinfo {year} {2003})}\BibitemShut
  {NoStop}%
\bibitem [{\citenamefont {Frustaglia}\ and\ \citenamefont
  {Richter}(2004)}]{Frustaglia2004}%
  \BibitemOpen
  \bibfield  {author} {\bibinfo {author} {\bibfnamefont {D.}~\bibnamefont
  {Frustaglia}}\ and\ \bibinfo {author} {\bibfnamefont {K.}~\bibnamefont
  {Richter}},\ }\href {\doibase 10.1103/PhysRevB.69.235310} {\bibfield
  {journal} {\bibinfo  {journal} {Phys. Rev. B}\ }\textbf {\bibinfo {volume}
  {69}},\ \bibinfo {pages} {235310} (\bibinfo {year} {2004})}\BibitemShut
  {NoStop}%
\bibitem [{\citenamefont {Nitta}\ and\ \citenamefont
  {Bergsten}(2007)}]{Nitta2007}%
  \BibitemOpen
  \bibfield  {author} {\bibinfo {author} {\bibfnamefont {J.}~\bibnamefont
  {Nitta}}\ and\ \bibinfo {author} {\bibfnamefont {T.}~\bibnamefont
  {Bergsten}},\ }\href {\doibase 10.1109/TED.2007.894370} {\bibfield  {journal}
  {\bibinfo  {journal} {IEEE Trans. Electron Devices}\ }\textbf {\bibinfo
  {volume} {54}},\ \bibinfo {pages} {955} (\bibinfo {year} {2007})}\BibitemShut
  {NoStop}%
\bibitem [{\citenamefont {Nagasawa}\ \emph {et~al.}(2013)\citenamefont
  {Nagasawa}, \citenamefont {Frustaglia}, \citenamefont {Saarikoski},
  \citenamefont {Richter},\ and\ \citenamefont {Nitta}}]{Nagasawa2013}%
  \BibitemOpen
  \bibfield  {author} {\bibinfo {author} {\bibfnamefont {F.}~\bibnamefont
  {Nagasawa}}, \bibinfo {author} {\bibfnamefont {D.}~\bibnamefont
  {Frustaglia}}, \bibinfo {author} {\bibfnamefont {H.}~\bibnamefont
  {Saarikoski}}, \bibinfo {author} {\bibfnamefont {K.}~\bibnamefont {Richter}},
  \ and\ \bibinfo {author} {\bibfnamefont {J.}~\bibnamefont {Nitta}},\ }\href
  {\doibase https://doi.org/10.1038/ncomms3526} {\bibfield  {journal} {\bibinfo
   {journal} {Nat. Commun.}\ }\textbf {\bibinfo {volume} {4}},\ \bibinfo
  {pages} {2526} (\bibinfo {year} {2013})}\BibitemShut {NoStop}%
\bibitem [{\citenamefont {Nagasawa}\ \emph {et~al.}(2018)\citenamefont
  {Nagasawa}, \citenamefont {Reynoso}, \citenamefont {Baltan\'as},
  \citenamefont {Frustaglia}, \citenamefont {Saarikoski},\ and\ \citenamefont
  {Nitta}}]{Nagasawa2018}%
  \BibitemOpen
  \bibfield  {author} {\bibinfo {author} {\bibfnamefont {F.}~\bibnamefont
  {Nagasawa}}, \bibinfo {author} {\bibfnamefont {A.~A.}\ \bibnamefont
  {Reynoso}}, \bibinfo {author} {\bibfnamefont {J.~P.}\ \bibnamefont
  {Baltan\'as}}, \bibinfo {author} {\bibfnamefont {D.}~\bibnamefont
  {Frustaglia}}, \bibinfo {author} {\bibfnamefont {H.}~\bibnamefont
  {Saarikoski}}, \ and\ \bibinfo {author} {\bibfnamefont {J.}~\bibnamefont
  {Nitta}},\ }\href {\doibase 10.1103/PhysRevB.98.245301} {\bibfield  {journal}
  {\bibinfo  {journal} {Phys. Rev. B}\ }\textbf {\bibinfo {volume} {98}},\
  \bibinfo {pages} {245301} (\bibinfo {year} {2018})}\BibitemShut {NoStop}%
\bibitem [{\citenamefont {Bernevig}\ \emph {et~al.}(2006)\citenamefont
  {Bernevig}, \citenamefont {Orenstein},\ and\ \citenamefont
  {Zhang}}]{Bernevig2006}%
  \BibitemOpen
  \bibfield  {author} {\bibinfo {author} {\bibfnamefont {B.~A.}\ \bibnamefont
  {Bernevig}}, \bibinfo {author} {\bibfnamefont {J.}~\bibnamefont {Orenstein}},
  \ and\ \bibinfo {author} {\bibfnamefont {S.-C.}\ \bibnamefont {Zhang}},\
  }\href {\doibase 10.1103/PhysRevLett.97.236601} {\bibfield  {journal}
  {\bibinfo  {journal} {Phys. Rev. Lett.}\ }\textbf {\bibinfo {volume} {97}},\
  \bibinfo {pages} {236601} (\bibinfo {year} {2006})}\BibitemShut {NoStop}%
\bibitem [{\citenamefont {Weber}\ \emph {et~al.}(2007)\citenamefont {Weber},
  \citenamefont {Orenstein}, \citenamefont {Bernevig}, \citenamefont {Zhang},
  \citenamefont {Stephens},\ and\ \citenamefont {Awschalom}}]{Weber2007}%
  \BibitemOpen
  \bibfield  {author} {\bibinfo {author} {\bibfnamefont {C.~P.}\ \bibnamefont
  {Weber}}, \bibinfo {author} {\bibfnamefont {J.}~\bibnamefont {Orenstein}},
  \bibinfo {author} {\bibfnamefont {B.~A.}\ \bibnamefont {Bernevig}}, \bibinfo
  {author} {\bibfnamefont {S.-C.}\ \bibnamefont {Zhang}}, \bibinfo {author}
  {\bibfnamefont {J.}~\bibnamefont {Stephens}}, \ and\ \bibinfo {author}
  {\bibfnamefont {D.~D.}\ \bibnamefont {Awschalom}},\ }\href {\doibase
  10.1103/PhysRevLett.98.076604} {\bibfield  {journal} {\bibinfo  {journal}
  {Phys. Rev. Lett.}\ }\textbf {\bibinfo {volume} {98}},\ \bibinfo {pages}
  {076604} (\bibinfo {year} {2007})}\BibitemShut {NoStop}%
\bibitem [{\citenamefont {Koralek}\ \emph {et~al.}(2009)\citenamefont
  {Koralek}, \citenamefont {Weber}, \citenamefont {Orenstein}, \citenamefont
  {Bernevig}, \citenamefont {Zhang}, \citenamefont {Mack},\ and\ \citenamefont
  {Awschalom}}]{Koralek2009}%
  \BibitemOpen
  \bibfield  {author} {\bibinfo {author} {\bibfnamefont {J.~D.}\ \bibnamefont
  {Koralek}}, \bibinfo {author} {\bibfnamefont {C.~P.}\ \bibnamefont {Weber}},
  \bibinfo {author} {\bibfnamefont {J.}~\bibnamefont {Orenstein}}, \bibinfo
  {author} {\bibfnamefont {B.~A.}\ \bibnamefont {Bernevig}}, \bibinfo {author}
  {\bibfnamefont {S.-C.}\ \bibnamefont {Zhang}}, \bibinfo {author}
  {\bibfnamefont {S.}~\bibnamefont {Mack}}, \ and\ \bibinfo {author}
  {\bibfnamefont {D.~D.}\ \bibnamefont {Awschalom}},\ }\href {\doibase
  https://doi.org/10.1038/nature07871} {\bibfield  {journal} {\bibinfo
  {journal} {Nat.}\ }\textbf {\bibinfo {volume} {458}},\ \bibinfo {pages} {610}
  (\bibinfo {year} {2009})}\BibitemShut {NoStop}%
\bibitem [{\citenamefont {Sasaki}\ \emph {et~al.}(2014)\citenamefont {Sasaki},
  \citenamefont {Nonaka}, \citenamefont {Kunihashi}, \citenamefont {Kohda},
  \citenamefont {Bauernfeind}, \citenamefont {Dollinger}, \citenamefont
  {Richter},\ and\ \citenamefont {Nitta}}]{Sasaki2014}%
  \BibitemOpen
  \bibfield  {author} {\bibinfo {author} {\bibfnamefont {A.}~\bibnamefont
  {Sasaki}}, \bibinfo {author} {\bibfnamefont {S.}~\bibnamefont {Nonaka}},
  \bibinfo {author} {\bibfnamefont {Y.}~\bibnamefont {Kunihashi}}, \bibinfo
  {author} {\bibfnamefont {M.}~\bibnamefont {Kohda}}, \bibinfo {author}
  {\bibfnamefont {T.}~\bibnamefont {Bauernfeind}}, \bibinfo {author}
  {\bibfnamefont {T.}~\bibnamefont {Dollinger}}, \bibinfo {author}
  {\bibfnamefont {K.}~\bibnamefont {Richter}}, \ and\ \bibinfo {author}
  {\bibfnamefont {J.}~\bibnamefont {Nitta}},\ }\href {\doibase
  https://doi.org/10.1038/nnano.2014.128} {\bibfield  {journal} {\bibinfo
  {journal} {Nat. Nanotechnol.}\ }\textbf {\bibinfo {volume} {9}},\ \bibinfo
  {pages} {703} (\bibinfo {year} {2014})}\BibitemShut {NoStop}%
\bibitem [{\citenamefont {Schliemann}(2017)}]{Schliemann2017}%
  \BibitemOpen
  \bibfield  {author} {\bibinfo {author} {\bibfnamefont {J.}~\bibnamefont
  {Schliemann}},\ }\href {\doibase 10.1103/RevModPhys.89.011001} {\bibfield
  {journal} {\bibinfo  {journal} {Rev. Mod. Phys.}\ }\textbf {\bibinfo {volume}
  {89}},\ \bibinfo {pages} {011001} (\bibinfo {year} {2017})}\BibitemShut
  {NoStop}%
\bibitem [{\citenamefont {Iizasa}\ \emph {et~al.}(2020)\citenamefont {Iizasa},
  \citenamefont {Kohda}, \citenamefont {Z\"ulicke}, \citenamefont {Nitta},\
  and\ \citenamefont {Kammermeier}}]{Iizasa2020}%
  \BibitemOpen
  \bibfield  {author} {\bibinfo {author} {\bibfnamefont {D.}~\bibnamefont
  {Iizasa}}, \bibinfo {author} {\bibfnamefont {M.}~\bibnamefont {Kohda}},
  \bibinfo {author} {\bibfnamefont {U.}~\bibnamefont {Z\"ulicke}}, \bibinfo
  {author} {\bibfnamefont {J.}~\bibnamefont {Nitta}}, \ and\ \bibinfo {author}
  {\bibfnamefont {M.}~\bibnamefont {Kammermeier}},\ }\href {\doibase
  10.1103/PhysRevB.101.245417} {\bibfield  {journal} {\bibinfo  {journal}
  {Phys. Rev. B}\ }\textbf {\bibinfo {volume} {101}},\ \bibinfo {pages}
  {245417} (\bibinfo {year} {2020})}\BibitemShut {NoStop}%
\bibitem [{\citenamefont {Zhao}\ \emph {et~al.}(2020)\citenamefont {Zhao},
  \citenamefont {Nakamura}, \citenamefont {Arras}, \citenamefont {Paillard},
  \citenamefont {Chen}, \citenamefont {Gosteau}, \citenamefont {Li},
  \citenamefont {Yang},\ and\ \citenamefont {Bellaiche}}]{Zhao2020}%
  \BibitemOpen
  \bibfield  {author} {\bibinfo {author} {\bibfnamefont {H.~J.}\ \bibnamefont
  {Zhao}}, \bibinfo {author} {\bibfnamefont {H.}~\bibnamefont {Nakamura}},
  \bibinfo {author} {\bibfnamefont {R.}~\bibnamefont {Arras}}, \bibinfo
  {author} {\bibfnamefont {C.}~\bibnamefont {Paillard}}, \bibinfo {author}
  {\bibfnamefont {P.}~\bibnamefont {Chen}}, \bibinfo {author} {\bibfnamefont
  {J.}~\bibnamefont {Gosteau}}, \bibinfo {author} {\bibfnamefont
  {X.}~\bibnamefont {Li}}, \bibinfo {author} {\bibfnamefont {Y.}~\bibnamefont
  {Yang}}, \ and\ \bibinfo {author} {\bibfnamefont {L.}~\bibnamefont
  {Bellaiche}},\ }\href {\doibase 10.1103/PhysRevLett.125.216405} {\bibfield
  {journal} {\bibinfo  {journal} {Phys. Rev. Lett.}\ }\textbf {\bibinfo
  {volume} {125}},\ \bibinfo {pages} {216405} (\bibinfo {year}
  {2020})}\BibitemShut {NoStop}%
\bibitem [{\citenamefont {Nakayama}\ \emph {et~al.}(2016)\citenamefont
  {Nakayama}, \citenamefont {Kanno}, \citenamefont {An}, \citenamefont
  {Tashiro}, \citenamefont {Haku}, \citenamefont {Nomura},\ and\ \citenamefont
  {Ando}}]{Nakayama2016}%
  \BibitemOpen
  \bibfield  {author} {\bibinfo {author} {\bibfnamefont {H.}~\bibnamefont
  {Nakayama}}, \bibinfo {author} {\bibfnamefont {Y.}~\bibnamefont {Kanno}},
  \bibinfo {author} {\bibfnamefont {H.}~\bibnamefont {An}}, \bibinfo {author}
  {\bibfnamefont {T.}~\bibnamefont {Tashiro}}, \bibinfo {author} {\bibfnamefont
  {S.}~\bibnamefont {Haku}}, \bibinfo {author} {\bibfnamefont {A.}~\bibnamefont
  {Nomura}}, \ and\ \bibinfo {author} {\bibfnamefont {K.}~\bibnamefont
  {Ando}},\ }\href {\doibase 10.1103/PhysRevLett.117.116602} {\bibfield
  {journal} {\bibinfo  {journal} {Phys. Rev. Lett.}\ }\textbf {\bibinfo
  {volume} {117}},\ \bibinfo {pages} {116602} (\bibinfo {year}
  {2016})}\BibitemShut {NoStop}%
\bibitem [{\citenamefont {S{\'a}nchez}\ \emph {et~al.}(2013)\citenamefont
  {S{\'a}nchez}, \citenamefont {Vila}, \citenamefont {Desfonds}, \citenamefont
  {Gambarelli}, \citenamefont {Attan{\'e}}, \citenamefont {De~Teresa},
  \citenamefont {Mag{\'e}n},\ and\ \citenamefont {Fert}}]{Sanchez2013}%
  \BibitemOpen
  \bibfield  {author} {\bibinfo {author} {\bibfnamefont {J.~C.~R.}\
  \bibnamefont {S{\'a}nchez}}, \bibinfo {author} {\bibfnamefont
  {L.}~\bibnamefont {Vila}}, \bibinfo {author} {\bibfnamefont {G.}~\bibnamefont
  {Desfonds}}, \bibinfo {author} {\bibfnamefont {S.}~\bibnamefont
  {Gambarelli}}, \bibinfo {author} {\bibfnamefont {J.~P.}\ \bibnamefont
  {Attan{\'e}}}, \bibinfo {author} {\bibfnamefont {J.~M.}\ \bibnamefont
  {De~Teresa}}, \bibinfo {author} {\bibfnamefont {C.}~\bibnamefont
  {Mag{\'e}n}}, \ and\ \bibinfo {author} {\bibfnamefont {A.}~\bibnamefont
  {Fert}},\ }\href {\doibase https://doi.org/10.1038/ncomms3944} {\bibfield
  {journal} {\bibinfo  {journal} {Nat. Commun.}\ }\textbf {\bibinfo {volume}
  {4}},\ \bibinfo {pages} {2944} (\bibinfo {year} {2013})}\BibitemShut
  {NoStop}%
\bibitem [{\citenamefont {Ghiasi}\ \emph {et~al.}(2019)\citenamefont {Ghiasi},
  \citenamefont {Kaverzin}, \citenamefont {Blah},\ and\ \citenamefont {van
  Wees}}]{Ghiasi2019}%
  \BibitemOpen
  \bibfield  {author} {\bibinfo {author} {\bibfnamefont {T.~S.}\ \bibnamefont
  {Ghiasi}}, \bibinfo {author} {\bibfnamefont {A.~A.}\ \bibnamefont
  {Kaverzin}}, \bibinfo {author} {\bibfnamefont {P.~J.}\ \bibnamefont {Blah}},
  \ and\ \bibinfo {author} {\bibfnamefont {B.~J.}\ \bibnamefont {van Wees}},\
  }\href {\doibase https://doi.org/10.1021/acs.nanolett.9b01611} {\bibfield
  {journal} {\bibinfo  {journal} {Nano Lett.}\ }\textbf {\bibinfo {volume}
  {19}},\ \bibinfo {pages} {5959} (\bibinfo {year} {2019})}\BibitemShut
  {NoStop}%
\bibitem [{\citenamefont {Inoue}\ \emph {et~al.}(2016)\citenamefont {Inoue},
  \citenamefont {Bauer},\ and\ \citenamefont {Nomura}}]{Inoue2016}%
  \BibitemOpen
  \bibfield  {author} {\bibinfo {author} {\bibfnamefont {T.}~\bibnamefont
  {Inoue}}, \bibinfo {author} {\bibfnamefont {G.~E.~W.}\ \bibnamefont {Bauer}},
  \ and\ \bibinfo {author} {\bibfnamefont {K.}~\bibnamefont {Nomura}},\ }\href
  {\doibase 10.1103/PhysRevB.94.205428} {\bibfield  {journal} {\bibinfo
  {journal} {Phys. Rev. B}\ }\textbf {\bibinfo {volume} {94}},\ \bibinfo
  {pages} {205428} (\bibinfo {year} {2016})}\BibitemShut {NoStop}%
\bibitem [{\citenamefont {Lesne}\ \emph {et~al.}(2016)\citenamefont {Lesne},
  \citenamefont {Fu}, \citenamefont {Oyarzun}, \citenamefont
  {Rojas-S{\'a}nchez}, \citenamefont {Vaz}, \citenamefont {Naganuma},
  \citenamefont {Sicoli}, \citenamefont {Attan{\'e}}, \citenamefont {Jamet},
  \citenamefont {Jacquet}, \citenamefont {George}, \citenamefont
  {Barth{\'e}l{\'e}my}, \citenamefont {Jaffr{\`e}s}, \citenamefont {Fert},
  \citenamefont {Bibes},\ and\ \citenamefont {Vila}}]{Lesne2016}%
  \BibitemOpen
  \bibfield  {author} {\bibinfo {author} {\bibfnamefont {E.}~\bibnamefont
  {Lesne}}, \bibinfo {author} {\bibfnamefont {Y.}~\bibnamefont {Fu}}, \bibinfo
  {author} {\bibfnamefont {S.}~\bibnamefont {Oyarzun}}, \bibinfo {author}
  {\bibfnamefont {J.~C.}\ \bibnamefont {Rojas-S{\'a}nchez}}, \bibinfo {author}
  {\bibfnamefont {D.~C.}\ \bibnamefont {Vaz}}, \bibinfo {author} {\bibfnamefont
  {H.}~\bibnamefont {Naganuma}}, \bibinfo {author} {\bibfnamefont
  {G.}~\bibnamefont {Sicoli}}, \bibinfo {author} {\bibfnamefont {J.-P.}\
  \bibnamefont {Attan{\'e}}}, \bibinfo {author} {\bibfnamefont
  {M.}~\bibnamefont {Jamet}}, \bibinfo {author} {\bibfnamefont
  {E.}~\bibnamefont {Jacquet}}, \bibinfo {author} {\bibfnamefont {J.-M.}\
  \bibnamefont {George}}, \bibinfo {author} {\bibfnamefont {A.}~\bibnamefont
  {Barth{\'e}l{\'e}my}}, \bibinfo {author} {\bibfnamefont {H.}~\bibnamefont
  {Jaffr{\`e}s}}, \bibinfo {author} {\bibfnamefont {A.}~\bibnamefont {Fert}},
  \bibinfo {author} {\bibfnamefont {M.}~\bibnamefont {Bibes}}, \ and\ \bibinfo
  {author} {\bibfnamefont {L.}~\bibnamefont {Vila}},\ }\href {\doibase
  https://doi.org/10.1038/nmat4726} {\bibfield  {journal} {\bibinfo  {journal}
  {Nat. Mater.}\ }\textbf {\bibinfo {volume} {15}},\ \bibinfo {pages} {1261}
  (\bibinfo {year} {2016})}\BibitemShut {NoStop}%
\bibitem [{\citenamefont {Song}\ \emph {et~al.}(2017)\citenamefont {Song},
  \citenamefont {Zhang}, \citenamefont {Su}, \citenamefont {Yuan},
  \citenamefont {Chen}, \citenamefont {Xing}, \citenamefont {Shi},
  \citenamefont {Sun},\ and\ \citenamefont {Han}}]{Song2017}%
  \BibitemOpen
  \bibfield  {author} {\bibinfo {author} {\bibfnamefont {Q.}~\bibnamefont
  {Song}}, \bibinfo {author} {\bibfnamefont {H.}~\bibnamefont {Zhang}},
  \bibinfo {author} {\bibfnamefont {T.}~\bibnamefont {Su}}, \bibinfo {author}
  {\bibfnamefont {W.}~\bibnamefont {Yuan}}, \bibinfo {author} {\bibfnamefont
  {Y.}~\bibnamefont {Chen}}, \bibinfo {author} {\bibfnamefont {W.}~\bibnamefont
  {Xing}}, \bibinfo {author} {\bibfnamefont {J.}~\bibnamefont {Shi}}, \bibinfo
  {author} {\bibfnamefont {J.}~\bibnamefont {Sun}}, \ and\ \bibinfo {author}
  {\bibfnamefont {W.}~\bibnamefont {Han}},\ }\href {\doibase
  10.1126/sciadv.1602312} {\bibfield  {journal} {\bibinfo  {journal} {Sci.
  Adv.}\ }\textbf {\bibinfo {volume} {3}},\ \bibinfo {pages} {e1602312}
  (\bibinfo {year} {2017})}\BibitemShut {NoStop}%
\bibitem [{\citenamefont {Ando}\ \emph {et~al.}(2011)\citenamefont {Ando},
  \citenamefont {Takahashi}, \citenamefont {Ieda}, \citenamefont {Kurebayashi},
  \citenamefont {Trypiniotis}, \citenamefont {Barnes}, \citenamefont
  {Maekawa},\ and\ \citenamefont {Saitoh}}]{Ando2011}%
  \BibitemOpen
  \bibfield  {author} {\bibinfo {author} {\bibfnamefont {K.}~\bibnamefont
  {Ando}}, \bibinfo {author} {\bibfnamefont {S.}~\bibnamefont {Takahashi}},
  \bibinfo {author} {\bibfnamefont {J.}~\bibnamefont {Ieda}}, \bibinfo {author}
  {\bibfnamefont {H.}~\bibnamefont {Kurebayashi}}, \bibinfo {author}
  {\bibfnamefont {T.}~\bibnamefont {Trypiniotis}}, \bibinfo {author}
  {\bibfnamefont {C.~H.~W.}\ \bibnamefont {Barnes}}, \bibinfo {author}
  {\bibfnamefont {S.}~\bibnamefont {Maekawa}}, \ and\ \bibinfo {author}
  {\bibfnamefont {E.}~\bibnamefont {Saitoh}},\ }\href {\doibase
  https://doi.org/10.1038/nmat3052} {\bibfield  {journal} {\bibinfo  {journal}
  {Nat. Mater.}\ }\textbf {\bibinfo {volume} {10}},\ \bibinfo {pages} {655}
  (\bibinfo {year} {2011})}\BibitemShut {NoStop}%
\bibitem [{\citenamefont {Sadovnikov}\ \emph {et~al.}(2019)\citenamefont
  {Sadovnikov}, \citenamefont {Beginin}, \citenamefont {Sheshukova},
  \citenamefont {Sharaevskii}, \citenamefont {Stognij}, \citenamefont
  {Novitski}, \citenamefont {Sakharov}, \citenamefont {Khivintsev},\ and\
  \citenamefont {Nikitov}}]{Sadovnikov2019}%
  \BibitemOpen
  \bibfield  {author} {\bibinfo {author} {\bibfnamefont {A.~V.}\ \bibnamefont
  {Sadovnikov}}, \bibinfo {author} {\bibfnamefont {E.~N.}\ \bibnamefont
  {Beginin}}, \bibinfo {author} {\bibfnamefont {S.~E.}\ \bibnamefont
  {Sheshukova}}, \bibinfo {author} {\bibfnamefont {Y.~P.}\ \bibnamefont
  {Sharaevskii}}, \bibinfo {author} {\bibfnamefont {A.~I.}\ \bibnamefont
  {Stognij}}, \bibinfo {author} {\bibfnamefont {N.~N.}\ \bibnamefont
  {Novitski}}, \bibinfo {author} {\bibfnamefont {V.~K.}\ \bibnamefont
  {Sakharov}}, \bibinfo {author} {\bibfnamefont {Y.~V.}\ \bibnamefont
  {Khivintsev}}, \ and\ \bibinfo {author} {\bibfnamefont {S.~A.}\ \bibnamefont
  {Nikitov}},\ }\href {\doibase 10.1103/PhysRevB.99.054424} {\bibfield
  {journal} {\bibinfo  {journal} {Phys. Rev. B}\ }\textbf {\bibinfo {volume}
  {99}},\ \bibinfo {pages} {054424} (\bibinfo {year} {2019})}\BibitemShut
  {NoStop}%
\bibitem [{\citenamefont {Ohnuma}\ \emph {et~al.}(2014)\citenamefont {Ohnuma},
  \citenamefont {Adachi}, \citenamefont {Saitoh},\ and\ \citenamefont
  {Maekawa}}]{Ohnuma2014}%
  \BibitemOpen
  \bibfield  {author} {\bibinfo {author} {\bibfnamefont {Y.}~\bibnamefont
  {Ohnuma}}, \bibinfo {author} {\bibfnamefont {H.}~\bibnamefont {Adachi}},
  \bibinfo {author} {\bibfnamefont {E.}~\bibnamefont {Saitoh}}, \ and\ \bibinfo
  {author} {\bibfnamefont {S.}~\bibnamefont {Maekawa}},\ }\href {\doibase
  10.1103/PhysRevB.89.174417} {\bibfield  {journal} {\bibinfo  {journal} {Phys.
  Rev. B}\ }\textbf {\bibinfo {volume} {89}},\ \bibinfo {pages} {174417}
  (\bibinfo {year} {2014})}\BibitemShut {NoStop}%
\bibitem [{\citenamefont {Matsuo}\ \emph {et~al.}(2018)\citenamefont {Matsuo},
  \citenamefont {Ohnuma}, \citenamefont {Kato},\ and\ \citenamefont
  {Maekawa}}]{Matsuo2018}%
  \BibitemOpen
  \bibfield  {author} {\bibinfo {author} {\bibfnamefont {M.}~\bibnamefont
  {Matsuo}}, \bibinfo {author} {\bibfnamefont {Y.}~\bibnamefont {Ohnuma}},
  \bibinfo {author} {\bibfnamefont {T.}~\bibnamefont {Kato}}, \ and\ \bibinfo
  {author} {\bibfnamefont {S.}~\bibnamefont {Maekawa}},\ }\href {\doibase
  10.1103/PhysRevLett.120.037201} {\bibfield  {journal} {\bibinfo  {journal}
  {Phys. Rev. Lett.}\ }\textbf {\bibinfo {volume} {120}},\ \bibinfo {pages}
  {037201} (\bibinfo {year} {2018})}\BibitemShut {NoStop}%
\bibitem [{\citenamefont {Kato}\ \emph {et~al.}(2019)\citenamefont {Kato},
  \citenamefont {Ohnuma}, \citenamefont {Matsuo}, \citenamefont {Rech},
  \citenamefont {Jonckheere},\ and\ \citenamefont {Martin}}]{Kato2019}%
  \BibitemOpen
  \bibfield  {author} {\bibinfo {author} {\bibfnamefont {T.}~\bibnamefont
  {Kato}}, \bibinfo {author} {\bibfnamefont {Y.}~\bibnamefont {Ohnuma}},
  \bibinfo {author} {\bibfnamefont {M.}~\bibnamefont {Matsuo}}, \bibinfo
  {author} {\bibfnamefont {J.}~\bibnamefont {Rech}}, \bibinfo {author}
  {\bibfnamefont {T.}~\bibnamefont {Jonckheere}}, \ and\ \bibinfo {author}
  {\bibfnamefont {T.}~\bibnamefont {Martin}},\ }\href {\doibase
  10.1103/PhysRevB.99.144411} {\bibfield  {journal} {\bibinfo  {journal} {Phys.
  Rev. B}\ }\textbf {\bibinfo {volume} {99}},\ \bibinfo {pages} {144411}
  (\bibinfo {year} {2019})}\BibitemShut {NoStop}%
\bibitem [{\citenamefont {Kato}\ \emph {et~al.}(2020)\citenamefont {Kato},
  \citenamefont {Ohnuma},\ and\ \citenamefont {Matsuo}}]{Kato2020}%
  \BibitemOpen
  \bibfield  {author} {\bibinfo {author} {\bibfnamefont {T.}~\bibnamefont
  {Kato}}, \bibinfo {author} {\bibfnamefont {Y.}~\bibnamefont {Ohnuma}}, \ and\
  \bibinfo {author} {\bibfnamefont {M.}~\bibnamefont {Matsuo}},\ }\href
  {\doibase 10.1103/PhysRevB.102.094437} {\bibfield  {journal} {\bibinfo
  {journal} {Phys. Rev. B}\ }\textbf {\bibinfo {volume} {102}},\ \bibinfo
  {pages} {094437} (\bibinfo {year} {2020})}\BibitemShut {NoStop}%
\bibitem [{\citenamefont {Ominato}\ and\ \citenamefont
  {Matsuo}(2020)}]{Ominato2020a}%
  \BibitemOpen
  \bibfield  {author} {\bibinfo {author} {\bibfnamefont {Y.}~\bibnamefont
  {Ominato}}\ and\ \bibinfo {author} {\bibfnamefont {M.}~\bibnamefont
  {Matsuo}},\ }\href {\doibase https://doi.org/10.7566/JPSJ.89.053704}
  {\bibfield  {journal} {\bibinfo  {journal} {J. Phys. Soc. Jpn.}\ }\textbf
  {\bibinfo {volume} {89}},\ \bibinfo {pages} {053704} (\bibinfo {year}
  {2020})}\BibitemShut {NoStop}%
\bibitem [{\citenamefont {Ominato}\ \emph {et~al.}(2020)\citenamefont
  {Ominato}, \citenamefont {Fujimoto},\ and\ \citenamefont
  {Matsuo}}]{Ominato2020b}%
  \BibitemOpen
  \bibfield  {author} {\bibinfo {author} {\bibfnamefont {Y.}~\bibnamefont
  {Ominato}}, \bibinfo {author} {\bibfnamefont {J.}~\bibnamefont {Fujimoto}}, \
  and\ \bibinfo {author} {\bibfnamefont {M.}~\bibnamefont {Matsuo}},\ }\href
  {\doibase 10.1103/PhysRevLett.124.166803} {\bibfield  {journal} {\bibinfo
  {journal} {Phys. Rev. Lett.}\ }\textbf {\bibinfo {volume} {124}},\ \bibinfo
  {pages} {166803} (\bibinfo {year} {2020})}\BibitemShut {NoStop}%
\bibitem [{Note1()}]{Note1}%
  \BibitemOpen
  \bibinfo {note} {{In Fig.~\ref {fig:spin_texture}, the spin-splitting of the
  Fermi surface has not been shown explicitly; it is assumed to be much smaller
  than the Fermi wavenumber.}}\BibitemShut {Stop}%
\bibitem [{\citenamefont {Miller}\ \emph {et~al.}(2003)\citenamefont {Miller},
  \citenamefont {Zumb\"uhl}, \citenamefont {Marcus}, \citenamefont
  {Lyanda-Geller}, \citenamefont {Goldhaber-Gordon}, \citenamefont {Campman},\
  and\ \citenamefont {Gossard}}]{Miller2003}%
  \BibitemOpen
  \bibfield  {author} {\bibinfo {author} {\bibfnamefont {J.~B.}\ \bibnamefont
  {Miller}}, \bibinfo {author} {\bibfnamefont {D.~M.}\ \bibnamefont
  {Zumb\"uhl}}, \bibinfo {author} {\bibfnamefont {C.~M.}\ \bibnamefont
  {Marcus}}, \bibinfo {author} {\bibfnamefont {Y.~B.}\ \bibnamefont
  {Lyanda-Geller}}, \bibinfo {author} {\bibfnamefont {D.}~\bibnamefont
  {Goldhaber-Gordon}}, \bibinfo {author} {\bibfnamefont {K.}~\bibnamefont
  {Campman}}, \ and\ \bibinfo {author} {\bibfnamefont {A.~C.}\ \bibnamefont
  {Gossard}},\ }\href {\doibase 10.1103/PhysRevLett.90.076807} {\bibfield
  {journal} {\bibinfo  {journal} {Phys. Rev. Lett.}\ }\textbf {\bibinfo
  {volume} {90}},\ \bibinfo {pages} {076807} (\bibinfo {year}
  {2003})}\BibitemShut {NoStop}%
\bibitem [{\citenamefont {Grundler}(2000)}]{Grundler2000}%
  \BibitemOpen
  \bibfield  {author} {\bibinfo {author} {\bibfnamefont {D.}~\bibnamefont
  {Grundler}},\ }\href {\doibase 10.1103/PhysRevLett.84.6074} {\bibfield
  {journal} {\bibinfo  {journal} {Phys. Rev. Lett.}\ }\textbf {\bibinfo
  {volume} {84}},\ \bibinfo {pages} {6074} (\bibinfo {year}
  {2000})}\BibitemShut {NoStop}%
\bibitem [{\citenamefont {Sato}\ \emph {et~al.}(2001)\citenamefont {Sato},
  \citenamefont {Kita}, \citenamefont {Gozu},\ and\ \citenamefont
  {Yamada}}]{Sato2001}%
  \BibitemOpen
  \bibfield  {author} {\bibinfo {author} {\bibfnamefont {Y.}~\bibnamefont
  {Sato}}, \bibinfo {author} {\bibfnamefont {T.}~\bibnamefont {Kita}}, \bibinfo
  {author} {\bibfnamefont {S.}~\bibnamefont {Gozu}}, \ and\ \bibinfo {author}
  {\bibfnamefont {S.}~\bibnamefont {Yamada}},\ }\href
  {https://doi.org/10.1063/1.1362356} {\bibfield  {journal} {\bibinfo
  {journal} {J. Appl. Phys.}\ }\textbf {\bibinfo {volume} {89}},\ \bibinfo
  {pages} {8017} (\bibinfo {year} {2001})}\BibitemShut {NoStop}%
\bibitem [{\citenamefont {Meier}\ \emph {et~al.}(2007)\citenamefont {Meier},
  \citenamefont {Salis}, \citenamefont {Shorubalko}, \citenamefont {Gini},
  \citenamefont {Sch{\"{o}}n},\ and\ \citenamefont {Ensslin}}]{Meier2007}%
  \BibitemOpen
  \bibfield  {author} {\bibinfo {author} {\bibfnamefont {L.}~\bibnamefont
  {Meier}}, \bibinfo {author} {\bibfnamefont {G.}~\bibnamefont {Salis}},
  \bibinfo {author} {\bibfnamefont {I.}~\bibnamefont {Shorubalko}}, \bibinfo
  {author} {\bibfnamefont {E.}~\bibnamefont {Gini}}, \bibinfo {author}
  {\bibfnamefont {S.}~\bibnamefont {Sch{\"{o}}n}}, \ and\ \bibinfo {author}
  {\bibfnamefont {K.}~\bibnamefont {Ensslin}},\ }\href {\doibase
  10.1038/nphys675} {\bibfield  {journal} {\bibinfo  {journal} {Nat. Phys.}\
  }\textbf {\bibinfo {volume} {3}},\ \bibinfo {pages} {650} (\bibinfo {year}
  {2007})}\BibitemShut {NoStop}%
\end{thebibliography}%

\end{document}